\documentclass[12pt]{article}
\usepackage{exscale}
\usepackage{axodraw}
\begin{document}
%================
%---
%--- Rosetta+.tex
%---
%--- Look at ``Temporary change!!!!!!!!!!'' 
%---
%================
% General utility
%================
%
\newcommand{\nl}{\nonumber\\}
\newcommand{\nn}{\nonumber}
\newcommand{\ds}{\displaystyle}
\newcommand{\mpar}[1]{{\marginpar{\hbadness10000%
                      \sloppy\hfuzz10pt\boldmath\bf#1}}%
                      \typeout{marginpar: #1}\ignorespaces}
\def\mnew{\mpar{\hfil NEW \hfil}\ignorespaces}
\newcommand{\lpar}{\left(}                            % bracketing
\newcommand{\rpar}{\right)} 
\newcommand{\lrbr}{\left[}
\newcommand{\rrbr}{\right]}
\newcommand{\lcbr}{\left\{}
\newcommand{\rcbr}{\right\}} 
\newcommand{\rbrak}[1]{\lrbr#1\rrbr}
\newcommand{\bq}{\begin{equation}}                    % equationing
\newcommand{\eq}{\end{equation}}
\newcommand{\bqa}{\begin{eqnarray}}
\newcommand{\eqa}{\end{eqnarray}}
\newcommand{\ba}[1]{\begin{array}{#1}}
\newcommand{\ea}{\end{array}}
\newcommand{\ben}{\begin{enumerate}}
\newcommand{\een}{\end{enumerate}}
\newcommand{\bei}{\begin{itemize}}
\newcommand{\eei}{\end{itemize}}
\newcommand{\eqn}[1]{Eq.(\ref{#1})}
\newcommand{\eqns}[2]{Eqs.(\ref{#1})--(\ref{#2})}
\newcommand{\eqnss}[1]{Eqs.(\ref{#1})}
\newcommand{\eqnsc}[2]{Eqs.(\ref{#1}) and (\ref{#2})}
\newcommand{\eqnst}[3]{Eqs.(\ref{#1}), (\ref{#2}) and (\ref{#3})}
\newcommand{\eqnsf}[4]{Eqs.(\ref{#1}), (\ref{#2}), (\ref{#3}) and (\ref{#4})}
\newcommand{\eqnsv}[5]{Eqs.(\ref{#1}), (\ref{#2}), (\ref{#3}), (\ref{#4}) and (\ref{#5})}
\newcommand{\tbn}[1]{Tab.~\ref{#1}}
\newcommand{\tabn}[1]{Tab.~\ref{#1}}
\newcommand{\tbns}[2]{Tabs.~\ref{#1}--\ref{#2}}
\newcommand{\tabns}[2]{Tabs.~\ref{#1}--\ref{#2}}
\newcommand{\tbnsc}[2]{Tabs.~\ref{#1} and \ref{#2}}
\newcommand{\fig}[1]{Fig.~\ref{#1}}
\newcommand{\figs}[2]{Figs.~\ref{#1}--\ref{#2}}
\newcommand{\sect}[1]{Section~\ref{#1}}
\newcommand{\sects}[2]{Section~\ref{#1} and \ref{#2}}
\newcommand{\subsect}[1]{Subsection~\ref{#1}}
\newcommand{\appendx}[1]{Appendix~\ref{#1}}
\newcommand{\hsp}{\hspace{.5mm}}
\def\negs{\hspace{-0.26in}}
\def\negsh{\hspace{-0.13in}}
%
% Miscellanea of symbols:
%========================
%
\newcommand{\TeV}{\mathrm{TeV}}                     
\newcommand{\GeV}{\mathrm{GeV}}
\newcommand{\MeV}{\mathrm{MeV}}
\newcommand{\nb}{\mathrm{nb}}
\newcommand{\pb}{\mathrm{pb}}
\newcommand{\fb}{\mathrm{fb}}
\def\Re{\mathop{\operator@font Re}\nolimits}
\def\Im{\mathop{\operator@font Im}\nolimits}
\newcommand{\ord}[1]{{\cal O}\lpar#1\rpar}
\newcommand{\group}{SU(2)\otimes U(1)}
\newcommand{\ib}{i}
\newcommand{\asums}[1]{\sum_{#1}}
\newcommand{\asumt}[2]{\sum_{#1}^{#2}}
\newcommand{\asum}[3]{\sum_{#1=#2}^{#3}}
%
% Powers of 10:
%==============
%
\newcommand{\tmi}{\times 10^{-1}}
\newcommand{\tmii}{\times 10^{-2}}
\newcommand{\tmiii}{\times 10^{-3}}
\newcommand{\tmiv}{\times 10^{-4}}
\newcommand{\tmfv}{\times 10^{-5}}
\newcommand{\tmfvi}{\times 10^{-6}}
\newcommand{\tmfvii}{\times 10^{-7}}
\newcommand{\tmfviii}{\times 10^{-8}}
\newcommand{\tmfix}{\times 10^{-9}}
\newcommand{\tmfx}{\times 10^{-10}}
%
% Fields:
%========
%
\newcommand{\fer}{{\rm{fer}}}
\newcommand{\bos}{{\rm{bos}}}
\newcommand{\lep}{{l}}
\newcommand{\had}{{h}}
\newcommand{\gen}{\rm{g}}
\newcommand{\dbl}{\rm{d}}
\newcommand{\philone}{\phi}
\newcommand{\philoneb}{\phi_{0}}
\newcommand{\phiind}[1]{\phi_{#1}}
\newcommand{\gBi}[2]{B_{#1}^{#2}}
\newcommand{\gBn}[1]{B^{#1}}
%
% vector-bosons
%--------------
%
\newcommand{\ph}{\gamma}
\newcommand{\ab}{A}
\newcommand{\abp}{A'}
\newcommand{\abr}{A^r}
\newcommand{\abb}{A^{0}}
\newcommand{\abi}[1]{A_{#1}}
\newcommand{\abpi}[1]{A'_{#1}}
\newcommand{\abri}[1]{A^r_{#1}}
\newcommand{\abbi}[1]{A^{0}_{#1}}
\newcommand{\wb}{W}            
\newcommand{\wbi}[1]{W_{#1}}           
\newcommand{\wbp}{W^{+}}
\newcommand{\wbm}{W^{-}}
\newcommand{\wbpm}{W^{\pm}}
\newcommand{\wbpi}[1]{W^{+}_{#1}}
\newcommand{\wbmi}[1]{W^{-}_{#1}}
\newcommand{\wbpmi}[1]{W^{\pm}_{#1}}
\newcommand{\wbmpi}[1]{W^{\mp}_{#1}}
\newcommand{\wbli}[1]{W^{[+}_{#1}}
\newcommand{\wbri}[1]{W^{-]}_{#1}}
\newcommand{\zb}{Z}
\newcommand{\zbi}[1]{Z_{#1}}
\newcommand{\vb}{V}
\newcommand{\vbi}[1]{V_{#1}}      
\newcommand{\vbiv}[1]{V^{*}_{#1}}      
\newcommand{\Pb}{P}
\newcommand{\Sb}{S}
\newcommand{\Bb}{B}
%
% Higgs-Kibble ghosts
%--------------------
%
\newcommand{\vph}{\varphi}
\newcommand{\hk}{K}
\newcommand{\hKi}[1]{K_{#1}}
\newcommand{\hkg}{\phi}
\newcommand{\hkn}{\phi^{0}}                 
\newcommand{\hkp}{\phi^{+}}
\newcommand{\hkm}{\phi^{-}}
\newcommand{\hkpm}{\phi^{\pm}}
\newcommand{\hkmp}{\phi^{\mp}}
\newcommand{\hki}[1]{\phi^{#1}}
\newcommand{\hb}{H}
\newcommand{\hbi}[1]{H_{#1}}
\newcommand{\hkl}{\phi^{[+\cgfi\cgfi}}
\newcommand{\hkr}{\phi^{-]}}
%
% FP-ghosts
%----------
%
\newcommand{\fpx}{X}
\newcommand{\fpy}{Y}
\newcommand{\fpxp}{X^+}
\newcommand{\fpxm}{X^-}
\newcommand{\fpxpm}{X^{\pm}}
\newcommand{\fpxi}[1]{X^{#1}}
\newcommand{\fpyZ}{Y^{\ssZ}}
\newcommand{\fpyA}{Y^{\ssA}}
\newcommand{\fpyG}{Y^{\ssG}}
\newcommand{\fpyZA}{Y_{\ssZ,\ssA}}
\newcommand{\fpbxi}[1]{{\overline{X}}^{#1}}
\newcommand{\fpbyZ}{{\overline{Y}}^{\ssZ}}
\newcommand{\fpbyA}{{\overline{Y}}^{\ssA}}
\newcommand{\fpbyZA}{{\overline{Y}}^{\ssZ,\ssA}}
%
% Fermionic fields
%-----------------
%
\newcommand{\Flone}{F}
\newcommand{\fpsi}{\psi}
\newcommand{\fpsif}{\psi_f}
\newcommand{\fpsifp}{\psi'_f}
\newcommand{\fpsii}[1]{\psi^{#1}}
\newcommand{\fbpsif}{{\overline{\psi}_f}}
\newcommand{\fbpsifp}{{\overline{\psi}'_f}}
\newcommand{\fpsib}{\psi_{\sszero}}
\newcommand{\fpsir}{\psi^r}
\newcommand{\fpsiL}{\psi_{_L}}
\newcommand{\fpsiR}{\psi_{_R}}
\newcommand{\fpsiLi}[1]{\psi_{_L}^{#1}}
\newcommand{\fpsiRi}[1]{\psi_{_R}^{#1}}
\newcommand{\fpsiLbi}[1]{\psi_{_{0L}}^{#1}}
\newcommand{\fpsiRbi}[1]{\psi_{_{0R}}^{#1}}
\newcommand{\fpsiLR}{\psi_{_{L,R}}}
\newcommand{\fbpsi}{{\overline{\psi}}}
\newcommand{\fbpsii}[1]{{\overline{\psi}}^{#1}}
\newcommand{\fbpsir}{{\overline{\psi}}^r}
\newcommand{\fbpsiL}{{\overline{\psi}}_{_L}}
\newcommand{\fbpsiR}{{\overline{\psi}}_{_R}}
\newcommand{\fbpsiLi}[1]{\overline{\psi_{_L}}^{#1}}
\newcommand{\fbpsiRi}[1]{\overline{\psi_{_R}}^{#1}}
\newcommand{\FQED}[2]{F_{#1#2}}
\newcommand{\FQEDp}[2]{F'_{#1#2}}
\newcommand{\fe}{e}
\newcommand{\ff}{f}
\newcommand{\fep}{e^{+}}
\newcommand{\fem}{e^{-}}
\newcommand{\fepm}{e^{\pm}}
\newcommand{\fp}{f^{+}}
\newcommand{\fm}{f^{-}}
\newcommand{\fhp}{h^{+}}
\newcommand{\fhm}{h^{-}}
\newcommand{\fh}{h}
\newcommand{\flm}{\mu}
\newcommand{\flmp}{\mu^{+}}
\newcommand{\flmm}{\mu^{-}}
\newcommand{\fll}{l}
\newcommand{\fllp}{l^{+}}
\newcommand{\fllm}{l^{-}}
\newcommand{\flt}{\tau}
\newcommand{\fltp}{\tau^{+}}
\newcommand{\fltm}{\tau^{-}}
\newcommand{\fq}{q}
\newcommand{\fqi}[1]{\fq_{#1}}
\newcommand{\bfqi}[1]{\barq_{#1}}
\newcommand{\ffQ}{Q}
\newcommand{\fu}{u}
\newcommand{\fd}{d}
\newcommand{\fc}{c}
\newcommand{\fs}{s}
\newcommand{\fqp}{q'}
\newcommand{\fup}{u'}
\newcommand{\fdp}{d'}
\newcommand{\fcp}{c'}
\newcommand{\fsp}{s'}
\newcommand{\fdpp}{d''}
\newcommand{\ffi}[1]{f_{#1}}
\newcommand{\bffi}[1]{{\overline{f}}_{#1}}
\newcommand{\ffpi}[1]{f'_{#1}}
\newcommand{\bffpi}[1]{{\overline{f}}'_{#1}}
\newcommand{\ft}{t}
\newcommand{\ffb}{b}
\newcommand{\ffp}{f'}
\newcommand{\fft}{{\tilde{f}}}
\newcommand{\fl}{l}
\newcommand{\fli}[1]{\fl_{#1}}
\newcommand{\fnu}{\nu}
\newcommand{\fU}{U}
\newcommand{\fD}{D}
\newcommand{\fUc}{\overline{U}}
\newcommand{\fDc}{\overline{D}}
\newcommand{\fnul}{\nu_l}
\newcommand{\fnue}{\nu_e}
\newcommand{\fnum}{\nu_{\mu}}
\newcommand{\fnut}{\nu_{\tau}}
\newcommand{\fbe}{{\overline{e}}}
\newcommand{\fblm}{{\overline{\mu}}}
\newcommand{\fblt}{{\overline{\tau}}}
\newcommand{\fbu}{{\overline{u}}}
\newcommand{\fbd}{{\overline{d}}}
\newcommand{\fbf}{{\overline{f}}}
\newcommand{\fbfp}{{\overline{f}}'}
\newcommand{\fbl}{{\overline{l}}}
\newcommand{\fbnu}{{\overline{\nu}}}
\newcommand{\fbnul}{{\overline{\nu}}_{\fl}}
\newcommand{\fbnue}{{\overline{\nu}}_{\fe}}
\newcommand{\fbnum}{{\overline{\nu}}_{\flm}}
\newcommand{\fbnut}{{\overline{\nu}}_{\flt}}
\newcommand{\fuL}{u_{_L}}
\newcommand{\fdL}{d_{_L}}
\newcommand{\ffL}{f_{_L}}
\newcommand{\fbuL}{{\overline{u}}_{_L}}
\newcommand{\fbdL}{{\overline{d}}_{_L}}
\newcommand{\fbfL}{{\overline{f}}_{_L}}
\newcommand{\fuR}{u_{_R}}
\newcommand{\fdR}{d_{_R}}
\newcommand{\ffR}{f_{_R}}
\newcommand{\fbuR}{{\overline{u}}_{_R}}
\newcommand{\fbdR}{{\overline{d}}_{_R}}
\newcommand{\fbfR}{{\overline{f}}_{_R}}
%
% anti-fermions, GP's realization
%--------------------------------
%
\newcommand{\barf}{\overline f}                
\newcommand{\barl}{\overline l}
\newcommand{\barq}{\overline q}
\newcommand{\barqp}{\overline{q}'}
\newcommand{\barb}{\overline b}
\newcommand{\bart}{\overline t}
\newcommand{\barc}{\overline c}
\newcommand{\baru}{\overline u}
\newcommand{\bard}{\overline d}
\newcommand{\bars}{\overline s}
\newcommand{\barv}{\overline v}
\newcommand{\barnu}{\overline{\nu}}
\newcommand{\barne}{\overline{\nu}_{\fe}}
\newcommand{\barnm}{\overline{\nu}_{\flm}}
\newcommand{\barnt}{\overline{\nu}_{\flt}}
%
% gluon
%------
%
\newcommand{\glu}{g}
%
% (anti)proton
%-------------
%
\newcommand{\prot}{p}
\newcommand{\aprot}{{\bar{p}}}
\newcommand{\Nucln}{N}
%
% Vector resonances
%------------------
%
\newcommand{\tM}{{\tilde M}}
\newcommand{\tMs}{{\tilde M}^2}
\newcommand{\tW}{{\tilde \Gamma}}
\newcommand{\tWs}{{\tilde\Gamma}^2}
\newcommand{\fphi}{\phi}
\newcommand{\fJpsi}{J/\psi}
\newcommand{\fgpsi}{\psi}
\newcommand{\Glone}{\Gamma}
\newcommand{\Gloni}[1]{\Gamma_{#1}}
\newcommand{\Glones}{\Gamma^2}
\newcommand{\Glonec}{\Gamma^3}
\newcommand{\glone}{\gamma}
\newcommand{\glones}{\gamma^2}
\newcommand{\gloneq}{\gamma^4}
\newcommand{\gloni}[1]{\gamma_{#1}}
\newcommand{\glonis}[1]{\gamma^2_{#1}}
\newcommand{\Grest}[2]{\Gamma_{#1}^{#2}}
\newcommand{\grest}[2]{\gamma_{#1}^{#2}}
\newcommand{\resampl}{A_{_{\rm{R}}}}
\newcommand{\resasyi}[1]{{\cal{A}}_{#1}}
\newcommand{\sSrest}[1]{\sigma_{#1}}
\newcommand{\Srest}[2]{\sigma_{#1}\lpar{#2}\rpar}
\newcommand{\Gdist}[1]{{\cal{G}}\lpar{#1}\rpar}
\newcommand{\sGdist}{{\cal{G}}}
\newcommand{\Aarea}{A_{0}}
\newcommand{\Aareai}[1]{{\cal{A}}\lpar{#1}\rpar}
\newcommand{\sAarea}{{\cal{A}}}
\newcommand{\resolw}{\sigma_{\ssE}}
\newcommand{\resolws}{\sigma^2_{\ssE}}
\newcommand{\chizer}{\chi_{0}}
\newcommand{\ini}{\rm{in}}
\newcommand{\fin}{\rm{fin}}
\newcommand{\ifi}{\rm{if}}
\newcommand{\ipf}{\rm{i+f}}
\newcommand{\tot}{\rm{T}}
\newcommand{\FB}{\rm{FB}}
\newcommand{\Nn}{\rm{N}}
\newcommand{\Bac}{\rm{Q}}
\newcommand{\Res}{\rm{R}}
\newcommand{\Int}{\rm{I}}
\newcommand{\NRe}{\rm{NR}}
\newcommand{\ratoe}{\delta}
\newcommand{\ratoes}{\delta^2}
%
% QED-boxes
%----------
%
\newcommand{\Fbox}[2]{f^{\rm{box}}_{#1}\lpar{#2}\rpar}
\newcommand{\Dbox}[2]{\delta^{\rm{box}}_{#1}\lpar{#2}\rpar}
\newcommand{\Bbox}[3]{{\cal{B}}_{#1}^{#2}\lpar{#3}\rpar}
%
% Masses:
%========
%
\newcommand{\phm}{\lambda}
\newcommand{\phms}{\lambda^2}
\newcommand{\mV}{M_{_V}}
\newcommand{\mw}{M_{_W}}
\newcommand{\mX}{M_{_X}}
\newcommand{\mY}{M_{_Y}}
\newcommand{\LM}{M}
\newcommand{\mz}{M_{_Z}}
\newcommand{\bzm}{M_{_0}}
\newcommand{\mh}{M_{_H}}
\newcommand{\bhm}{M_{_{0H}}}
\newcommand{\mf}{m_f}
\newcommand{\mfp}{m_{f'}}
\newcommand{\mfh}{m_{h}}
\newcommand{\mt}{m_t}
\newcommand{\me}{m_e}
\newcommand{\mm}{m_{\mu}}
\newcommand{\mtau}{m_{\tau}}
\newcommand{\muq}{m_u}
\newcommand{\md}{m_d}
\newcommand{\muqp}{m'_u}
\newcommand{\mdqp}{m'_d}
\newcommand{\mc}{m_c}
\newcommand{\ms}{m_s}
\newcommand{\mb}{m_b}
\newcommand{\mup}{M_u}                              % pole masses
\newcommand{\mdp}{M_d}
\newcommand{\mcp}{M_c}
\newcommand{\msp}{M_s}
\newcommand{\mbp}{M_b}
%
% Masses squared, cubed and higher
%---------------------------------
%
\newcommand{\mls}{m^2_l}
\newcommand{\mVs}{M^2_{_V}}
\newcommand{\mVq}{M^4_{_V}}
\newcommand{\mws}{M^2_{_W}}
\newcommand{\mwc}{M^3_{_W}}
\newcommand{\LMs}{M^2}
\newcommand{\LMc}{M^3}
\newcommand{\mzs}{M^2_{_Z}}
\newcommand{\mzc}{M^3_{_Z}}
\newcommand{\bzms}{M^2_{_0}}
\newcommand{\bzmc}{M^3_{_0}}
\newcommand{\bhms}{M^2_{_{0H}}}
\newcommand{\mhs}{M^2_{_H}}
\newcommand{\mfs}{m^2_f}
\newcommand{\mfc}{m^3_f}
\newcommand{\mfps}{m^2_{f'}}
\newcommand{\mfhs}{m^2_{h}}
\newcommand{\mfpc}{m^3_{f'}}
\newcommand{\mts}{m^2_t}
\newcommand{\mes}{m^2_e}
\newcommand{\mms}{m^2_{\mu}}
\newcommand{\mmc}{m^3_{\mu}}
\newcommand{\mmfour}{m^4_{\mu}}
\newcommand{\mmf}{m^5_{\mu}}
\newcommand{\mmfive}{m^5_{\mu}}
\newcommand{\mmsix}{m^6_{\mu}}
\newcommand{\mtsix}{m^6_{\ft}}
\newcommand{\mminv}{\frac{1}{m_{\mu}}}
\newcommand{\mtaus}{m^2_{\tau}}
\newcommand{\mus}{m^2_u}
\newcommand{\mds}{m^2_d}
\newcommand{\muqps}{m'^2_u}
\newcommand{\mdqps}{m'^2_d}
\newcommand{\mcs}{m^2_c}
\newcommand{\mss}{m^2_s}
\newcommand{\mbs}{m^2_b}
\newcommand{\mups}{M^2_u}
\newcommand{\mdps}{M^2_d}
\newcommand{\mcps}{M^2_c}
\newcommand{\msps}{M^2_s}
\newcommand{\mbps}{M^2_b}
%
% Some ratios, special notations
%-------------------------------
%
\newcommand{\muf}{\mu_{\ff}}
\newcommand{\mufs}{\mu^2_{\ff}}
\newcommand{\mufq}{\mu^4_{\ff}}
\newcommand{\mufx}{\mu^6_{\ff}}
\newcommand{\muz}{\mu_{_{\zb}}}
\newcommand{\muw}{\mu_{_{\wb}}}
\newcommand{\mut}{\mu_{\ft}}
\newcommand{\muzs}{\mu^2_{_{\zb}}}
\newcommand{\muws}{\mu^2_{_{\wb}}}
\newcommand{\muts}{\mu^2_{\ft}}
\newcommand{\mubs}{\mu^2_{\ffb}}
\newcommand{\muSW}{\mu^2_{_{\wb}}}
\newcommand{\muwq}{\mu^4_{_{\wb}}}
\newcommand{\muwsx}{\mu^6_{_{\wb}}}
\newcommand{\muwms}{\mu^{-2}_{_{\wb}}}
\newcommand{\muhs}{\mu^2_{_{\hb}}}
\newcommand{\muhq}{\mu^4_{_{\hb}}}
\newcommand{\muhsx}{\mu^6_{_{\hb}}}
\newcommand{\mutq}{\mu^4_{_{\hb}}}   % bardinworry to be checked by grep 
\newcommand{\mutsx}{\mu^6_{_{\hb}}}  %     -/-
\newcommand{\muL}{\mu}
\newcommand{\muS}{\mu^2}
\newcommand{\muQ}{\mu^4}
\newcommand{\muizs}{\mu^2_{0}}
\newcommand{\muizq}{\mu^4_{0}}
\newcommand{\muis}{\mu^2_{1}}
\newcommand{\muiis}{\mu^2_{2}}
\newcommand{\muiiis}{\mu^2_{3}}
\newcommand{\muii}[1]{\mu_{#1}}
\newcommand{\muisi}[1]{\mu^2_{#1}}
\newcommand{\muiqi}[1]{\mu^4_{#1}}
\newcommand{\muixi}[1]{\mu^6_{#1}}
\newcommand{\zm}{z_m}
\newcommand{\mwzs}{M^2_{\ssW,\ssZ}}
\newcommand{\ri}[1]{r_{#1}}
\newcommand{\xw}{x_w}
\newcommand{\xws}{x^2_w}
\newcommand{\xwc}{x^3_w}
\newcommand{\xth}{x_t}
\newcommand{\xths}{x^2_t}
\newcommand{\xthc}{x^3_t}
\newcommand{\xthf}{x^4_t}
\newcommand{\xthv}{x^5_t}
\newcommand{\xthx}{x^6_t}
\newcommand{\xh}{x_h}
\newcommand{\xhs}{x^2_h}
\newcommand{\xhc}{x^3_h}
\newcommand{\Rl}{R_{\fl}}
\newcommand{\Rb}{R_{\ffb}}
\newcommand{\Rc}{R_{\fc}}
%
% Masses quartic
%---------------
%
\newcommand{\mwq}{M^4_{_\wb}}
\newcommand{\mwf}{M^4_{_\wb}}
\newcommand{\LMq}{M^4}
\newcommand{\mzq}{M^4_{_Z}}
\newcommand{\bzmq}{M^4_{_0}}
\newcommand{\mhq}{M^4_{_H}}
\newcommand{\mfq}{m^4_f}
\newcommand{\mfpq}{m^4_{f'}}
\newcommand{\mtq}{m^4_t}
\newcommand{\meq}{m^4_e}
\newcommand{\mmq}{m^4_{\mu}}
\newcommand{\mtauq}{m^4_{\tau}}
\newcommand{\muqq}{m^4_u}
\newcommand{\mdq}{m^4_d}
\newcommand{\mcq}{m^4_c}
\newcommand{\msq}{m^4_s}
\newcommand{\mbq}{m^4_b}
\newcommand{\mupq}{M^4_u}
\newcommand{\mdpq}{M^4_d}
\newcommand{\mcpq}{M^4_c}
\newcommand{\mspq}{M^4_s}
\newcommand{\mbpq}{M^4_b}
%
% Masses sixtupled
%-----------------
%
\newcommand{\mwx}{M^6_{_W}}
\newcommand{\mzx}{M^6_{_Z}}
\newcommand{\mfx}{m^6_f}
\newcommand{\mfpx}{m^6_{f'}}
\newcommand{\LMx}{M^6}
%
% More masses
%------------
%
\newcommand{\mer}{m_{er}}
\newcommand{\mlep}{m_l}
\newcommand{\mleps}{m^2_l}
\newcommand{\mzer}{m_0}
\newcommand{\mone}{m_1}
\newcommand{\mtwo}{m_2}
\newcommand{\mtre}{m_3}
\newcommand{\mfor}{m_4}
\newcommand{\mlone}{m}
\newcommand{\mloneb}{\bar{m}}
\newcommand{\mind}[1]{m_{#1}}
\newcommand{\mzers}{m^2_0}
\newcommand{\mones}{m^2_1}
\newcommand{\mtwos}{m^2_2}
\newcommand{\mtres}{m^2_3}
\newcommand{\mfors}{m^2_4}
\newcommand{\mlones}{m^2}
\newcommand{\minds}[1]{m^2_{#1}}
\newcommand{\mlonec}{m^3}
\newcommand{\moneq}{m^4_1}
\newcommand{\mtwoq}{m^4_2}
\newcommand{\mtreq}{m^4_3}
\newcommand{\mforq}{m^4_4}
\newcommand{\mloneq}{m^4}
\newcommand{\mindq}[1]{m^4_{#1}}
\newcommand{\mlonev}{m^5}
\newcommand{\mindv}[1]{m^5_{#1}}
\newcommand{\monex}{m^6_1}
\newcommand{\mtwox}{m^6_2}
\newcommand{\mtrex}{m^6_3}
\newcommand{\mforx}{m^6_4}
\newcommand{\mlonex}{m^6}
\newcommand{\mindx}[1]{m^6_{#1}}
\newcommand{\Mone}{M_1}
\newcommand{\Mtwo}{M_2}
\newcommand{\Mtre}{M_3}
\newcommand{\Mfor}{M_4}
\newcommand{\Mlone}{M}
\newcommand{\Mlonep}{M'}
\newcommand{\Miind}{M_i}
\newcommand{\Mind}[1]{M_{#1}}
\newcommand{\Minds}[1]{M^2_{#1}}
\newcommand{\Mindc}[1]{M^3_{#1}}
\newcommand{\Mindf}[1]{M^4_{#1}}
\newcommand{\Mones}{M^2_1}
\newcommand{\Mtwos}{M^2_2}
\newcommand{\Mtres}{M^2_3}
\newcommand{\Mfors}{M^2_4}
\newcommand{\Mlones}{M^2}
\newcommand{\Mloneps}{M'^2}
\newcommand{\Miinds}{M^2_i}
\newcommand{\Mlonec}{M^3}
\newcommand{\Monec}{M^3_1}
\newcommand{\Mtwoc}{M^3_2}
\newcommand{\Moneq}{M^4_1}
\newcommand{\Mtwoq}{M^4_2}
\newcommand{\Mtreq}{M^4_3}
\newcommand{\Mforq}{M^4_4}
\newcommand{\Mloneq}{M^4}
\newcommand{\Miindq}{M^4_i}
\newcommand{\Monex}{M^6_1}
\newcommand{\Mtwox}{M^6_2}
\newcommand{\Mtrex}{M^6_3}
\newcommand{\Mforx}{M^6_4}
\newcommand{\Mlonex}{M^6}
\newcommand{\Miindx}{M^6_i}
\newcommand{\meb}{m_{\sszero}}
\newcommand{\mebs}{m^2_{\sszero}}
%
% Pole masses again
%------------------
%
\newcommand{\Mq }{M_q  }
\newcommand{\MqS}{M^2_q}
\newcommand{\Ms }{M_s  }
\newcommand{\MsS}{M^2_s}
\newcommand{\Mc }{M_c  }
\newcommand{\McS}{M^2_c}
\newcommand{\Mb }{M_b  }
\newcommand{\MbS}{M^2_b}
\newcommand{\Mt }{M_t  }
\newcommand{\MtS}{M^2_t}
%
% Some quark masses
%------------------
%
\newcommand{\mq}{m_q}
\newcommand{\mqs}{m^2_q}
\newcommand{\mqS}{m^2_q}
\newcommand{\mqQ}{m^4_q}
\newcommand{\mqX}{m^6_q}
\newcommand{\mqp}{m'_q }
\newcommand{\mqpS}{m'^2_q}
\newcommand{\mqpQ}{m'^4_q}
%
% Some logs of mass ratios
%=========================
%
\newcommand{\lL}{l}
\newcommand{\lLi}[1]{l_{#1}}
\newcommand{\ls}{l^2}
\newcommand{\LL}{L}
\newcommand{\LcalL}{\cal{L}}
\newcommand{\LS}{L^2}
\newcommand{\LC}{L^3}
\newcommand{\LQ}{L^4}
\newcommand{\lw}{l_w}
\newcommand{\Lw}{L_w}
\newcommand{\Lws}{L^2_w}
\newcommand{\Lz}{L_z}
\newcommand{\Lzs}{L^2_z}
\newcommand{\Li}[1]{L_{#1}}
\newcommand{\Lis}[1]{L^2_{#1}}
\newcommand{\Lic}[1]{L^3_{#1}}
%
% Mandelstam variables
%=====================
%
\newcommand{\sman}{s}
\newcommand{\tman}{t}
\newcommand{\uman}{u}
\newcommand{\smani}[1]{s_{#1}}
\newcommand{\bsmani}[1]{{\bar{s}}_{#1}}
\newcommand{\smans}{s^2}
\newcommand{\tmans}{t^2}
\newcommand{\umans}{u^2}
\newcommand{\shat}{{\hat s}}
\newcommand{\that}{{\hat t}}
\newcommand{\uhat}{{\hat u}}
%
% More invariant variables
%-------------------------
%
\newcommand{\smanp}{s'}
\newcommand{\smanpi}[1]{s'_{#1}}
\newcommand{\tmanp}{t'}
\newcommand{\umanp}{u'}
\newcommand{\kappi}[1]{\kappa_{#1}}
\newcommand{\zetai}[1]{\zeta_{#1}}
%
% QED
%====
%
% Phase space and QED-varia
%--------------------------
%
\newcommand{\Phaspi}[1]{\Gamma_{#1}}
\newcommand{\rbetai}[1]{\beta_{#1}}
\newcommand{\ralphai}[1]{\alpha_{#1}}
\newcommand{\rbetais}[1]{\beta^2_{#1}}
\newcommand{\Lambdi}[1]{\Lambda_{#1}}
\newcommand{\Nomini}[1]{N_{#1}}
\newcommand{\smlone}{\frac{-\sman-\ib\ep}{\mlones}}
%
% Angles in bremsstrahlung
%-------------------------
%
\newcommand{\theti}[1]{\theta_{#1}}
\newcommand{\delti}[1]{\delta_{#1}}
\newcommand{\phigi}[1]{\phi_{#1}}
\newcommand{\acoli}[1]{\xi_{#1}}
\newcommand{\scats}{s}
\newcommand{\scatss}{s^2}
\newcommand{\scatsi}[1]{s_{#1}}
\newcommand{\scatsis}[1]{s^2_{#1}}
\newcommand{\scatst}[2]{s_{#1}^{#2}}
\newcommand{\scatc}{c}
\newcommand{\scatcs}{c^2}
\newcommand{\scatci}[1]{c_{#1}}
\newcommand{\scatcis}[1]{c^2_{#1}}
\newcommand{\scatct}[2]{c_{#1}^{#2}}
\newcommand{\angamt}[2]{\gamma_{#1}^{#2}}
\newcommand{\scatsin}{\sin\theta}
\newcommand{\scatsins}{\sin^2\theta}
\newcommand{\scatcos}{\cos\theta}
\newcommand{\scatcoss}{\cos^2\theta}
%
% More about bremsstrahlung
%--------------------------
%
\newcommand{\Regia}{{\cal{R}}}
\newcommand{\Iconi}[2]{{\cal{I}}_{#1}\lpar{#2}\rpar}
\newcommand{\sIcon}[1]{{\cal{I}}_{#1}}
\newcommand{\betaf}{\beta_{\ff}}
\newcommand{\betafs}{\beta^2_{\ff}}
\newcommand{\betafc}{\beta^3_{\ff}}
\newcommand{\Kfact}[2]{{\cal{K}}_{#1}\lpar{#2}\rpar}
%
% Structure and flux functions
%-----------------------------
%
\newcommand{\Struf}[4]{{\cal D}^{#1}_{#2}\lpar{#3;#4}\rpar}
\newcommand{\sStruf}[2]{{\cal D}^{#1}_{#2}}
\newcommand{\Fluxf}[2]{H\lpar{#1;#2}\rpar}
\newcommand{\Fluxfi}[4]{H_{#1}^{#2}\lpar{#3;#4}\rpar}
\newcommand{\sFluxf}{H}
\newcommand{\Bflux}[2]{{\cal{B}}_{#1}\lpar{#2}\rpar}
\newcommand{\bflux}[2]{{\cal{B}}_{#1}\lpar{#2}\rpar}
\newcommand{\Fluxd}[2]{D_{#1}\lpar{#2}\rpar}
\newcommand{\fluxd}[2]{C_{#1}\lpar{#2}\rpar}
\newcommand{\Fluxh}[4]{{\cal{H}}_{#1}^{#2}\lpar{#3;#4}\rpar}
\newcommand{\Sluxh}[4]{{\cal{S}}_{#1}^{#2}\lpar{#3;#4}\rpar}
\newcommand{\Fluxhb}[4]{{\overline{{\cal{H}}}}_{#1}^{#2}\lpar{#3;#4}\rpar}
\newcommand{\sFluxhb}{{\overline{{\cal{H}}}}}
\newcommand{\Sluxhb}[4]{{\overline{{\cal{S}}}}_{#1}^{#2}\lpar{#3;#4}\rpar}
\newcommand{\sSluxhb}[2]{{\overline{{\cal{S}}}}_{#1}^{#2}}
\newcommand{\fluxh}[4]{h_{#1}^{#2}\lpar{#3;#4}\rpar}
\newcommand{\fluxhs}[3]{h_{#1}^{#2}\lpar{#3}\rpar}
\newcommand{\sfluxhs}[2]{h_{#1}^{#2}}
\newcommand{\fluxhb}[4]{{\overline{h}}_{#1}^{#2}\lpar{#3;#4}\rpar}
\newcommand{\Strufd}[2]{D\lpar{#1;#2}\rpar}
%
% Mass and momenta squared ratios
%================================
%
\newcommand{\rMQ}[1]{r^2_{#1}}
\newcommand{\rMQs}[1]{r^4_{#1}}
\newcommand{\hf}{h_{\ff}}
\newcommand{\rf}{w_{\ff}}
\newcommand{\zf}{z_{\ff}}
\newcommand{\rfs}{w^2_{\ff}}
\newcommand{\zfs}{z^2_{\ff}}
\newcommand{\rfc}{w^3_{\ff}}
\newcommand{\zfc}{z^3_{\ff}}
\newcommand{\df}{d_{\ff}}
\newcommand{\rfp}{w_{\ffp}}
\newcommand{\rfps}{w^2_{\ffp}}
\newcommand{\rfpc}{w^3_{\ffp}}
\newcommand{\Lrfp}{L_{w}}
\newcommand{\rt}{w_{\ft}}
\newcommand{\rts}{w^2_{\ft}}
\newcommand{\rb}{w_{\ffb}}
\newcommand{\rbs}{w^2_{\ffb}}
\newcommand{\dt}{d_{\ft}}
\newcommand{\dts}{d^2_{\ft}}
\newcommand{\rh}{r_{h}}
\newcommand{\Lnrt}{\ln{\rt}}
\newcommand{\Rw}{R_{_{\wb}}}
\newcommand{\Rws}{R^2_{_{\wb}}}
\newcommand{\Rz}{R_{_{\zb}}}
\newcommand{\Rzp}{R^{+}_{_{\zb}}}
\newcommand{\Rzm}{R^{-}_{_{\zb}}}
\newcommand{\Rzs}{R^2_{_{\zb}}}
\newcommand{\Rzc}{R^3_{_{\zb}}}
\newcommand{\Rv}{R_{_{\vb}}}
\newcommand{\rhw}{w_h}
\newcommand{\rhz}{z_h}
\newcommand{\rhws}{w^2_h}
\newcommand{\rhzs}{z^2_h}
%
% More ratios
%------------
%
\newcommand{\vqrato}{z}
\newcommand{\vqrats}{w}
\newcommand{\vqratq}{w^2}
\newcommand{\seyrat}{z}
\newcommand{\sexrat}{w}
\newcommand{\sexrats}{w^2}
\newcommand{\sehrat}{h}
\newcommand{\sewrat}{w}
\newcommand{\sezrat}{z}
\newcommand{\zetav}{\zeta}
\newcommand{\zetavi}[1]{\zeta_{#1}}
\newcommand{\bpo}{\beta^2}
\newcommand{\bpos}{\beta^4}
\newcommand{\bpt}{{\tilde\beta}^2}
\newcommand{\lap}{\kappa}
\newcommand{\hw}{w_h}
\newcommand{\hz}{z_h}
%
% Couplings
%==========
%
\newcommand{\ec}{e}
\newcommand{\ecs}{e^2}
\newcommand{\ect}{e^3}
\newcommand{\ecq}{e^4}
\newcommand{\ecb}{e_{\sszero}}
\newcommand{\ecbs}{e^2_{_0}}
\newcommand{\ecbq}{e^4_{_0}}
\newcommand{\eci}[1]{e_{#1}}
\newcommand{\ecis}[1]{e^2_{#1}}
\newcommand{\hate}{{\hat e}}
\newcommand{\gss}{g_{_S}}
\newcommand{\gsss}{g^2_{_S}}
\newcommand{\gssb}{g^2_{_{S_0}}}
\newcommand{\als}{\alpha_{_S}}
\newcommand{\as}{a_{_S}}
\newcommand{\ass}{a^2_{_S}}
\newcommand{\gf}{G_{\ssF}}
\newcommand{\gfs}{G^2_{\ssF}}
\newcommand{\gb}{g} 
\newcommand{\gbi}[1]{g_{#1}}
\newcommand{\gbb}{g_{0}}
\newcommand{\gbs}{g^2}
\newcommand{\gbc}{g^3}
\newcommand{\gbf}{g^4}
\newcommand{\gpb}{g'}
\newcommand{\gpbs}{g'^2}
\newcommand{\vc}[1]{v_{#1}}
\newcommand{\ac}[1]{a_{#1}}
\newcommand{\vcc}[1]{v^*_{#1}}
\newcommand{\acc}[1]{a^*_{#1}}
\newcommand{\hatv}[1]{{\hat v}_{#1}}
\newcommand{\vcs}[1]{v^2_{#1}}
\newcommand{\acs}[1]{a^2_{#1}}
\newcommand{\gcv}[1]{g^{#1}_{\ssV}}
\newcommand{\gca}[1]{g^{#1}_{\ssA}}
\newcommand{\gcp}[1]{g^{+}_{#1}}
\newcommand{\gcm}[1]{g^{-}_{#1}}
\newcommand{\gcpm}[1]{g^{\pm}_{#1}}
\newcommand{\vci}[2]{v^{#2}_{#1}}
\newcommand{\aci}[2]{a^{#2}_{#1}}
\newcommand{\vceff}[1]{v^{#1}_{\rm{eff}}}
\newcommand{\hvc}[1]{\hat{v}_{#1}}
\newcommand{\hvcs}[1]{\hat{v}^2_{#1}}
\newcommand{\Vc}[1]{V_{#1}}
\newcommand{\Ac}[1]{A_{#1}}
\newcommand{\Vcs}[1]{V^2_{#1}}
\newcommand{\Acs}[1]{A^2_{#1}}
\newcommand{\vpa}[2]{\sigma_{#1}^{#2}}
\newcommand{\vma}[2]{\delta_{#1}^{#2}}
\newcommand{\vfw}{\sigma^{a}_{\ff}}
\newcommand{\vfpw}{\sigma^{a}_{\ffp}}
\newcommand{\vfwi}[1]{\sigma^{a}_{#1}}
\newcommand{\vfwsi}[1]{\lpar\sigma^{a}_{#1}\rpar^2}
\newcommand{\vvfw}{v^{a}_{\ff}}
\newcommand{\vvew}{v^{a}_{\fe}}
\newcommand{\vzm}{v^{-}_{\ssZ}}
\newcommand{\vzp}{v^{+}_{\ssZ}}
\newcommand{\vzpm}{v^{\pm}_{\ssZ}}
\newcommand{\vzmp}{v^{\mp}_{\ssZ}}
\newcommand{\vam}{v^{-}_{\ssA}}
\newcommand{\vap}{v^{+}_{\ssA}}
\newcommand{\vapm}{v^{\pm}_{\ssA}}
%-------------------------------------------------> g,G - couplings 
\newcommand{\gv}{g_{_V}}
\newcommand{\ga}{g_{_A}}
\newcommand{\gve}{g^{\fe}_{_{V}}}
\newcommand{\gae}{g^{\fe}_{_{A}}}
\newcommand{\gvf}{g^{\ff}_{_{V}}}
\newcommand{\gaf}{g^{\ff}_{_{A}}}
\newcommand{\gva}{g_{_{V,A}}}
\newcommand{\gvae}{g^{\fe}_{_{V,A}}}
\newcommand{\gvaf}{g^{\ff}_{_{V,A}}}
\newcommand{\sGv}{{\cal{G}}_{_V}}
\newcommand{\cGa}{{\cal{G}}^{*}_{_A}}
\newcommand{\cGv}{{\cal{G}}^{*}_{_V}}
\newcommand{\sGa}{{\cal{G}}_{_A}}
\newcommand{\Gvf}{{\cal{G}}^{\ff}_{_{V}}}
\newcommand{\Gaf}{{\cal{G}}^{\ff}_{_{A}}}
\newcommand{\Gvaf}{{\cal{G}}^{\ff}_{_{V,A}}}
\newcommand{\Gve}{{\cal{G}}^{\fe}_{_{V}}}
\newcommand{\Gae}{{\cal{G}}^{\fe}_{_{A}}}
\newcommand{\Gvae}{{\cal{G}}^{\fe}_{_{V,A}}}
\newcommand{\gvl}{g^{\fl}_{_{V}}}
\newcommand{\gal}{g^{\fl}_{_{A}}}
\newcommand{\gval}{g^{\fl}_{_{V,A}}}
\newcommand{\gvb}{g^{\ffb}_{_{V}}}
\newcommand{\gab}{g^{\ffb}_{_{A}}}
\newcommand{\fvf}{F_{_V}^{\ff}}
\newcommand{\faf}{F_{_A}^{\ff}}
\newcommand{\fvl}{F_{_V}^{\fl}}
\newcommand{\fal}{F_{_A}^{\fl}}
\newcommand{\corat}{\kappa}
\newcommand{\corats}{\kappa^2}
%
% Deltology-rhoology-kappaology
%==============================
%                              
\newcommand{\dr}{\Delta r}
\newcommand{\drl}{\Delta r_{_L}}
\newcommand{\drh}{\Delta{\hat r}}
\newcommand{\drhw}{\Delta{\hat r}_{_W}}
\newcommand{\rhou}{\rho_{_U}}
\newcommand{\rhoz}{\rho_{_\zb}}
\newcommand{\rZ}{\rho_{_\zb}}
\newcommand{\rhob}{\rho_{_0}}
\newcommand{\rZf}{\rho^{\ff}_{_\zb}}
\newcommand{\rhoe}{\rho_{\fe}}
\newcommand{\rhof}{\rho_{\ff}}
\newcommand{\rhoi}[1]{\rho_{#1}}
\newcommand{\kZf}{\kappa^{\ff}_{_\zb}}
\newcommand{\rWf}{\rho^{\ff}_{_\wb}}
\newcommand{\brWf}{{\bar{\rho}}^{\ff}_{_\wb}}
\newcommand{\rHf}{\rho^{\ff}_{_\hb}}
\newcommand{\brHf}{{\bar{\rho}}^{\ff}_{_\hb}}
\newcommand{\rhoR}{\rho^{\ssR}_{_{\zb}}}
\newcommand{\hatrh}{{\hat\rho}}
\newcommand{\ku}{\kappa_{\ssU}}
\newcommand{\rZdf}[1]{\rho^{#1}_{_\zb}}
\newcommand{\kZdf}[1]{\kappa^{#1}_{_\zb}}
\newcommand{\rdfL}[1]{\rho^{#1}_{_L}}
\newcommand{\kdfL}[1]{\kappa^{#1}_{_L}}
\newcommand{\rdfR}[1]{\rho^{#1}_{\rm{rem}}}
\newcommand{\kdfR}[1]{\kappa^{#1}_{\rm{rem}}}
\newcommand{\bark}{\overline\kappa}
%
% Weak mixing angles
%===================
%
\newcommand{\stw}{s_{\theta}}             % bare, Lagrangian parameters
\newcommand{\ctw}{c_{\theta}}
\newcommand{\stws}{s_{\theta}^2}
\newcommand{\stwc}{s_{\theta}^3}
\newcommand{\stwf}{s_{\theta}^4}
\newcommand{\stwx}{s_{\theta}^6}
\newcommand{\ctws}{c_{\theta}^2}
\newcommand{\ctwc}{c_{\theta}^3}
\newcommand{\ctwf}{c_{\theta}^4}
\newcommand{\ctwx}{c_{\theta}^6}
\newcommand{\stwfiv}{s_{\theta}^5}
\newcommand{\ctwfiv}{c_{\theta}^5}
\newcommand{\stwsix}{s_{\theta}^6}
\newcommand{\ctwsix}{c_{\theta}^6}
%
% on-shell sines
%---------------
%
\newcommand{\siw}{s_{_W}}           
\newcommand{\cow}{c_{_W}}
\newcommand{\siws}{s^2_{_W}}
\newcommand{\cows}{c^2_{_W}}
\newcommand{\siwc}{s^3_{_W}}
\newcommand{\cowc}{c^3_{_W}}
\newcommand{\siwf}{s^4_{_W}}
\newcommand{\cowf}{c^4_{_W}}
\newcommand{\siwx}{s^6_{_W}}
\newcommand{\cowx}{c^6_{_W}}
\newcommand{\sons}{s_{_W}}
\newcommand{\sonss}{s^2_{_W}}
\newcommand{\cons}{c_{_W}}
\newcommand{\cooss}{c^2_{_W}}
%
% effective and other weak mixing angles
%---------------------------------------
%
\newcommand{\szs}{{\overline s}^2}
\newcommand{\szq}{{\overline s}^4}
\newcommand{\czs}{{\overline c}^2}
\newcommand{\sbs}{s_{_0}^2}
\newcommand{\cbs}{c_{_0}^2}
\newcommand{\dss}{\Delta s^2}
\newcommand{\snes}{s_{\nu e}^2}
\newcommand{\cnes}{c_{\nu e}^2}
\newcommand{\shs}{{\hat s}^2}
\newcommand{\chs}{{\hat c}^2}
\newcommand{\chl}{{\hat c}}
\newcommand{\seffs}{s^2_{\rm{eff}}}
\newcommand{\seffsf}[1]{\sin^2\theta^{#1}_{\rm{eff}}}
\newcommand{\sress}{s^2_{\rm res}}                
\newcommand{\sR}{s_{_R}}
\newcommand{\sRs}{s^2_{_R}}
\newcommand{\ctwe}{c_{\theta}^6}
\newcommand{\sany}{s}
\newcommand{\cany}{c}
\newcommand{\sanys}{s^2}
\newcommand{\canys}{c^2}
%
% Spinology etc.
%===============
%
\newcommand{\sip}{u}                             % incoming particle
\newcommand{\siap}{{\bar{v}}}                    %    "     anti-p
\newcommand{\sop}{{\bar{u}}}                     % outgoing p
\newcommand{\soap}{v}                            %    "     anti-p
\newcommand{\ip}[1]{u\lpar{#1}\rpar}             % incoming particle
\newcommand{\iap}[1]{{\bar{v}}\lpar{#1}\rpar}    %    "     anti-p
\newcommand{\op}[1]{{\bar{u}}\lpar{#1}\rpar}     % outgoing p
\newcommand{\oap}[1]{v\lpar{#1}\rpar}            %    "     anti-p
%
% With polarization
%------------------
%
\newcommand{\ipp}[2]{u\lpar{#1,#2}\rpar}         % incoming particle
\newcommand{\ipap}[2]{{\bar v}\lpar{#1,#2}\rpar} %    "     anti-p
\newcommand{\opp}[2]{{\bar u}\lpar{#1,#2}\rpar}  % outgoing p
\newcommand{\opap}[2]{v\lpar{#1,#2}\rpar}        %    "     anti-p
\newcommand{\upspi}[1]{u\lpar{#1}\rpar}
\newcommand{\vpspi}[1]{v\lpar{#1}\rpar}
\newcommand{\wpspi}[1]{w\lpar{#1}\rpar}
\newcommand{\ubpspi}[1]{{\bar{u}}\lpar{#1}\rpar}
\newcommand{\vbpspi}[1]{{\bar{v}}\lpar{#1}\rpar}
\newcommand{\wbpspi}[1]{{\bar{w}}\lpar{#1}\rpar}
\newcommand{\udpspi}[1]{u^{\dagger}\lpar{#1}\rpar}
\newcommand{\vdpspi}[1]{v^{\dagger}\lpar{#1}\rpar}
\newcommand{\wdpspi}[1]{w^{\dagger}\lpar{#1}\rpar}
\newcommand{\Ubilin}[1]{U\lpar{#1}\rpar}
\newcommand{\Vbilin}[1]{V\lpar{#1}\rpar}
\newcommand{\Xbilin}[1]{X\lpar{#1}\rpar}
\newcommand{\Ybilin}[1]{Y\lpar{#1}\rpar}
\newcommand{\up}[2]{u_{#1}\lpar #2\rpar}
\newcommand{\vp}[2]{v_{#1}\lpar #2\rpar}
\newcommand{\ubp}[2]{{\overline u}_{#1}\lpar #2\rpar}
\newcommand{\vbp}[2]{{\overline v}_{#1}\lpar #2\rpar}
\newcommand{\Pje}[1]{\frac{1}{2}\lpar 1 + #1\,\gfd\rpar}
\newcommand{\Pj}[1]{\Pi_{#1}}
\newcommand{\trace}{\mbox{Tr}}
%
% polarization operators and related things
%==========================================
%
\newcommand{\Poper}[2]{P_{#1}\lpar{#2}\rpar}
\newcommand{\Loper}[2]{\Lambda_{#1}\lpar{#2}\rpar}
\newcommand{\proj}[3]{P_{#1}\lpar{#2,#3}\rpar}
\newcommand{\sproj}[1]{P_{#1}}
\newcommand{\Nden}[3]{N_{#1}^{#2}\lpar{#3}\rpar}
\newcommand{\sNden}[1]{N_{#1}}
\newcommand{\nden}[2]{n_{#1}^{#2}}
%
% Wave functions
%===============
%
\newcommand{\vwf}[2]{e_{#1}\lpar#2\rpar}             % vector wave funct.
\newcommand{\vwfb}[2]{{\overline e}_{#1}\lpar#2\rpar}
\newcommand{\pwf}[2]{\epsilon_{#1}\lpar#2\rpar}      % photon wave funct.
\newcommand{\sla}[1]{/\!\!\!#1}
\newcommand{\slac}[1]{/\!\!\!\!#1}
%
% Momenta
%========
%
\newcommand{\iemom}{p_{_-}}                    % 2f incoming momenta
\newcommand{\ipmom}{p_{_+}}
\newcommand{\oemom}{q_{_-}}                    % 2f outgoing momenta
\newcommand{\opmom}{q_{_+}}
%
% Scalar product of two momenta
%==============================
%
\newcommand{\spro}[2]{{#1}\cdot{#2}}
%
% gammas
%=======
%
\newcommand{\gfour}{\gamma_4}                    
\newcommand{\gfd}{\gamma_5}                    
\newcommand{\gap}{\lpar 1+\gamma_5\rpar}
\newcommand{\gam}{\lpar 1-\gamma_5\rpar}
\newcommand{\gdp}{\gamma_+}
\newcommand{\gdm}{\gamma_-}
\newcommand{\gdpm}{\gamma_{\pm}}
\newcommand{\gad}{\gamma}
\newcommand{\gapm}{\lpar 1\pm\gamma_5\rpar}
\newcommand{\gadi}[1]{\gamma_{#1}}
\newcommand{\gadu}[1]{\gamma_{#1}}
\newcommand{\gaduc}[1]{\gamma^{*}_{#1}}
\newcommand{\gaduh}[1]{\gamma^{+}_{#1}}
\newcommand{\gadut}[1]{\gamma^{\ssT}_{#1}}
\newcommand{\gapu}[1]{\gamma^{#1}}
\newcommand{\sigd}[2]{\sigma_{#1#2}}
\newcommand{\sumsp}{\overline{\sum_{\mbox{\tiny{spins}}}}}
%
% Special functions & integrals
%==============================
%
\newcommand{\li}[2]{\mathrm{Li}_{#1}\lpar\displaystyle{#2}\rpar} % polylog
\newcommand{\sli}[1]{\mathrm{Li}_{#1}} % polylog without arg
\newcommand{\etaf}[2]{\eta\lpar#1,#2\rpar}
\newcommand{\lkall}[3]{\lambda\lpar#1,#2,#3\rpar}       % Kallen's lambda
\newcommand{\slkall}[3]{\lambda^{1/2}\lpar#1,#2,#3\rpar}%\sqrt
\newcommand{\segam}{\Gamma}                             % Euler's Gamma
\newcommand{\egam}[1]{\Gamma\lpar#1\rpar}               % Euler's Gamma
\newcommand{\egams}[1]{\Gamma^2\lpar#1\rpar}            % Euler's Gamma squared
\newcommand{\ebe}[2]{B\lpar#1,#2\rpar}                  % Euler's beta
\newcommand{\ddel}[1]{\delta\lpar#1\rpar}               % Dirac's delta
\newcommand{\ddeln}[1]{\delta^{(n)}\lpar#1\rpar}        % Dirac's delta in n-dim
\newcommand{\drii}[2]{\delta_{#1#2}}                    % Kronecker's delta
\newcommand{\driv}[4]{\delta_{#1#2#3#4}}                % gen.   "
\newcommand{\intmomi}[2]{\int\,d^{#1}#2}
\newcommand{\intmomii}[3]{\int\,d^{#1}#2\,\int\,d^{#1}#3}
\newcommand{\intfx}[1]{\int_{\scriptstyle 0}^{\scriptstyle 1}\,d#1}
\newcommand{\intfxy}[2]{\int_{\scriptstyle 0}^{\scriptstyle 1}\,d#1\,
                        \int_{\scriptstyle 0}^{\scriptstyle #1}\,d#2}
\newcommand{\intfxyz}[3]{\int_{\scriptstyle 0}^{\scriptstyle 1}\,d#1\,
                         \int_{\scriptstyle 0}^{\scriptstyle #1}\,d#2\,
                         \int_{\scriptstyle 0}^{\scriptstyle #2}\,d#3}
\newcommand{\Beta}[2]{{\rm{B}}\lpar #1,#2\rpar}
\newcommand{\sBeta}{\rm{B}}
\newcommand{\sign}[1]{{\rm{sign}}\lpar{#1}\rpar}
%
% Z widths
%=========
%
\newcommand{\gn}{\Gamma_{\nu}}
\newcommand{\gel}{\Gamma_{\fe}}
\newcommand{\gmu}{\Gamma_{\mu}}
\newcommand{\gff}{\Gamma_{\ff}}
\newcommand{\gt}{\Gamma_{\tau}}
\newcommand{\gl}{\Gamma_{\fl}}
\newcommand{\gq}{\Gamma_{\fq}}
\newcommand{\gu}{\Gamma_{\fu}}
\newcommand{\gd}{\Gamma_{\fd}}
\newcommand{\gc}{\Gamma_{\fc}}
\newcommand{\gs}{\Gamma_{\fs}}
\newcommand{\gbq}{\Gamma_{\ffb}}
\newcommand{\gz}{\Gamma_{_{\zb}}}
\newcommand{\gw}{\Gamma_{_{\wb}}}
\newcommand{\gh}{\Gamma_{\had}}
\newcommand{\ghb}{\Gamma_{_{\hb}}}
\newcommand{\gi}{\Gamma_{\rm{inv}}}
\newcommand{\gzs}{\Gamma^2_{_{\zb}}}
%
% Quantum numbers
%================
%
\newcommand{\tcie}{I^{(3)}_{\fe}}
\newcommand{\tcim}{I^{(3)}_{\flm}}
\newcommand{\tcif}{I^{(3)}_{\ff}}
\newcommand{\tciq}{I^{(3)}_{\fq}}
\newcommand{\tcib}{I^{(3)}_{\ffb}}
\newcommand{\tcih}{I^{(3)}_h}
\newcommand{\tcii}{I^{(3)}_i}
\newcommand{\tcift}{I^{(3)}_{\tilde f}}
\newcommand{\tcifp}{I^{(3)}_{f'}}
\newcommand{\wispt}[1]{I^{(3)}_{#1}}
\newcommand{\ql}{Q_l}
\newcommand{\qe}{Q_e}
\newcommand{\qu}{Q_u}
\newcommand{\qd}{Q_d}
\newcommand{\qb}{Q_b}
\newcommand{\qt}{Q_t}
\newcommand{\qup}{Q'_u}
\newcommand{\qdp}{Q'_d}
\newcommand{\qmu}{Q_{\mu}}
\newcommand{\qes}{Q^2_e}
\newcommand{\qec}{Q^3_e}
\newcommand{\qus}{Q^2_u}
\newcommand{\qds}{Q^2_d}
\newcommand{\qbs}{Q^2_b}
\newcommand{\qts}{Q^2_t}
\newcommand{\qbc}{Q^3_b}
\newcommand{\qf}{Q_f}
\newcommand{\qfs}{Q^2_f}
\newcommand{\qfc}{Q^3_f}
\newcommand{\qff}{Q^4_f}
\newcommand{\qep}{Q_{e'}}
\newcommand{\qfp}{Q_{f'}}
\newcommand{\qfps}{Q^2_{f'}}
\newcommand{\qfpc}{Q^3_{f'}}
\newcommand{\qq}{Q_q}
\newcommand{\qqs}{Q^2_q}
\newcommand{\qi}{Q_i}
\newcommand{\qis}{Q^2_i}
\newcommand{\qj}{Q_j}
\newcommand{\qjs}{Q^2_j}
\newcommand{\QW}{Q_{_\wb}}
\newcommand{\QWs}{Q^2_{_\wb}}
\newcommand{\Qd}{Q_d}
\newcommand{\Qds}{Q^2_d}
\newcommand{\Qu}{Q_u}
\newcommand{\Qus}{Q^2_u}
\newcommand{\vi}{v_i}
\newcommand{\vis}{v^2_i}
\newcommand{\ai}{a_i}
\newcommand{\ais}{a^2_i}
%
% Self-energies
%==============
%
\newcommand{\piv}{\Pi_{_V}}
\newcommand{\pia}{\Pi_{_A}}
\newcommand{\piva}{\Pi_{_{V,A}}}
\newcommand{\pivi}[1]{\Pi^{({#1})}_{_V}}
\newcommand{\piai}[1]{\Pi^{({#1})}_{_A}}
\newcommand{\pivai}[1]{\Pi^{({#1})}_{_{V,A}}}
\newcommand{\pih}{{\hat\Pi}}
\newcommand{\sgh}{{\hat\Sigma}}
\newcommand{\Pgg}{\Pi_{\ph\ph}}
\newcommand{\Ptg}{\Pi_{_{3Q}}}
\newcommand{\Ptt}{\Pi_{_{33}}}
\newcommand{\Pzg}{\Pi_{_{\zb\ab}}}
\newcommand{\Pzga}[2]{\Pi^{#1}_{_{\zb\ab}}\lpar#2\rpar}
\newcommand{\Pf}{\Pi^{\ssF}}
\newcommand{\Sgg}{\Sigma_{_{\ab\ab}}}
\newcommand{\Szg}{\Sigma_{_{\zb\ab}}}
\newcommand{\SVV}{\Sigma_{_{\vb\vb}}}
\newcommand{\USvv}{{\hat\Sigma}_{_{\vb\vb}}}
\newcommand{\Sww}{\Sigma_{_{\wb\wb}}}
\newcommand{\Swwg}{\Sigma^{_G}_{_{\wb\wb}}}
\newcommand{\Szz}{\Sigma_{_{\zb\zb}}}
\newcommand{\Shh}{\Sigma_{_{\hb\hb}}}
\newcommand{\Spzz}{\Sigma'_{_{\zb\zb}}}
\newcommand{\Stg}{\Sigma_{_{3Q}}}
\newcommand{\Stt}{\Sigma_{_{33}}}
\newcommand{\bSww}{{\overline\Sigma}_{_{WW}}}
\newcommand{\bStg}{{\overline\Sigma}_{_{3Q}}}
\newcommand{\bStt}{{\overline\Sigma}_{_{33}}}
\newcommand{\Sssn}{\Sigma_{_{\hkn\hkn}}}
\newcommand{\Sssc}{\Sigma_{_{\phi\phi}}}
\newcommand{\Szn}{\Sigma_{_{\zb\hkn}}}
\newcommand{\Swc}{\Sigma_{_{\wb\hkg}}}
\newcommand{\mix}[2]{{\cal{M}}^{#1}\lpar{#2}\rpar}
\newcommand{\bmix}[2]{\Pi^{{#1},\ssF}_{_{\zb\ab}}\lpar{#2}\rpar}
\newcommand{\hPgg}[2]{{\hat{\Pi}^{{#1},\ssF}}_{\ph\ph}\lpar{#2}\rpar}
\newcommand{\hmix}[2]{{\hat{\Pi}^{{#1},\ssF}}_{_{\zb\ab}}\lpar{#2}\rpar}
\newcommand{\Dz}[2]{{\cal{D}}_{_{\zb}}^{#1}\lpar{#2}\rpar}
\newcommand{\bDz}[2]{{\cal{D}}^{{#1},\ssF}_{_{\zb}}\lpar{#2}\rpar}
\newcommand{\hDz}[2]{{\hat{\cal{D}}}^{{#1},\ssF}_{_{\zb}}\lpar{#2}\rpar}
\newcommand{\Szzd}[2]{\Sigma'^{#1}_{_{\zb\zb}}\lpar{#2}\rpar}
\newcommand{\Swwd}[2]{\Sigma'^{#1}_{_{\wb\wb}}\lpar{#2}\rpar}
\newcommand{\Shhd}[2]{\Sigma'^{#1}_{_{\hb\hb}}\lpar{#2}\rpar}
\newcommand{\ZFren}[2]{{\cal{Z}}^{#1}\lpar{#2}\rpar}
\newcommand{\WFren}[2]{{\cal{W}}^{#1}\lpar{#2}\rpar}
\newcommand{\HFren}[2]{{\cal{H}}^{#1}\lpar{#2}\rpar}
\newcommand{\WI}{\cal{W}}
%
% QCD varia
%==========
%
\newcommand{\cf}{c_f}
\newcommand{\Cf}{C_{_F}}
\newcommand{\Nf}{N_f}
\newcommand{\Nc}{N_c}
\newcommand{\Ncs}{N^2_c}
\newcommand{\nf }{n_f}
\newcommand{\nfs}{n^2_f}
\newcommand{\nfc}{n^3_f}
\newcommand{\MSB}{\overline{MS}}
\newcommand{\LMSB}{\Lambda_{\overline{\mathrm{MS}}}}
\newcommand{\LMSBp}{\Lambda'_{\overline{\mathrm{MS}}}}
\newcommand{\LMSBS}{\Lambda^2_{\overline{\mathrm{MS}}}}
\newcommand{\LMSBv }{\mbox{$\Lambda^{(5)}_{\overline{\mathrm{MS}}}$}}
\newcommand{\LMSBvS}{\mbox{$\left(\Lambda^{(5)}_{\overline{\mathrm{MS}}}\right)^2$}}
\newcommand{\LMSBt }{\mbox{$\Lambda^{(3)}_{\overline{\mathrm{MS}}}$}}
\newcommand{\LMSBtS}{\mbox{$\left(\Lambda^{(3)}_{\overline{\mathrm{MS}}}\right)^2$}}
\newcommand{\LMSBf }{\mbox{$\Lambda^{(4)}_{\overline{\mathrm{MS}}}$}}
\newcommand{\LMSBfS}{\mbox{$\left(\Lambda^{(4)}_{\overline{\mathrm{MS}}}\right)^2$}}
\newcommand{\LMSBn }{\mbox{$\Lambda^{(\nf)}_{\overline{\mathrm{MS}}}$}}
\newcommand{\LMSBnS}{\mbox{$\left(\Lambda^{(\nf)}_{\overline{\mathrm{MS}}}\right)^2$}}
\newcommand{\LMSBnml }{\mbox{$\Lambda^{(\nf-1)}_{\overline{\mathrm{MS}}}$}}
\newcommand{\LMSBnmlS}{\mbox{$\left(\Lambda^{(\nf-1)}_{\overline{\mathrm{MS}}}\right)^2$}}
\newcommand{\Bnf}{\lpar\nf \rpar}
\newcommand{\Bnfm}{\lpar\nf-1 \rpar}
\newcommand{\LuM}{L_{_M}}
\newcommand{\bef}{\beta_{\ff}}
\newcommand{\befs}{\beta^2_{\ff}}
\newcommand{\befc}{\beta^3_{f}}
\newcommand{\alsp}{\alpha'_{_S}}
\newcommand{\api}{\displaystyle \frac{\als(s)}{\pi}}
\newcommand{\alss}{\alpha^2_{_S}}
\newcommand{\ztwo}{\zeta(2)}
\newcommand{\ztri}{\zeta(3)}
\newcommand{\zfor}{\zeta(4)}
\newcommand{\zfiv}{\zeta(5)}
\newcommand{\bi}[1]{b_{#1}}
\newcommand{\ci}[1]{c_{#1}}
\newcommand{\Ci}[1]{C_{#1}}
\newcommand{\bip}[1]{b'_{#1}}
\newcommand{\cip}[1]{c'_{#1}}
%
% Numerical factors
%==================
%
\newcommand{\osps}{16\,\pi^2}
\newcommand{\srt}{\sqrt{2}}
\newcommand{\ospsi}{\displaystyle{\frac{i}{16\,\pi^2}}}
%
% 2f processes
%=============
%
\newcommand{\tfpromu}{\mbox{$e^+e^-\to \mu^+\mu^-$}}
\newcommand{\tfprotau}{\mbox{$e^+e^-\to \tau^+\tau^-$}}
\newcommand{\tfproe}{\mbox{$e^+e^-\to e^+e^-$}}
\newcommand{\tfpronu}{\mbox{$e^+e^-\to \barnu\nu$}}
\newcommand{\tfproqq}{\mbox{$e^+e^-\to \barq q$}}
\newcommand{\tfprohad}{\mbox{$e^+e^-\to\,$} hadrons}
%
% brems. processes
%-----------------
%
\newcommand{\bpromu}{\mbox{$e^+e^-\to \mu^+\mu^-\ph$}}
\newcommand{\bprotau}{\mbox{$e^+e^-\to \tau^+\tau^-\ph$}}
\newcommand{\bproe}{\mbox{$e^+e^-\to e^+e^-\ph$}}
\newcommand{\bpronu}{\mbox{$e^+e^-\to \barnu\nu\ph$}}
\newcommand{\bproqq}{\mbox{$e^+e^-\to \barq q \ph$}}
%
% 2b processes
%-------------
%
\newcommand{\tbprow} {\mbox{$e^+e^-\to \wbp \wbm $}}
\newcommand{\tbproz} {\mbox{$e^+e^-\to \zb  \zb  $}}
\newcommand{\tbproh} {\mbox{$e^+e^-\to \zb  \hb  $}}
\newcommand{\tbprozg}{\mbox{$e^+e^-\to \zb  \ph  $}}
\newcommand{\tbprog} {\mbox{$e^+e^-\to \ph  \ph  $}}
%
% Line-style for propagators
%===========================
%
\newcommand{\Fermionline}[1]{
\vcenter{\hbox{
  \begin{picture}(60,20)(0,{#1})
  \SetScale{2.}
    \ArrowLine(0,5)(30,5)
  \end{picture}}}
}
\newcommand{\AntiFermionline}[1]{
\vcenter{\hbox{
  \begin{picture}(60,20)(0,{#1})
  \SetScale{2.}
    \ArrowLine(30,5)(0,5)
  \end{picture}}}
}
\newcommand{\Photonline}[1]{
\vcenter{\hbox{
  \begin{picture}(60,20)(0,{#1})
  \SetScale{2.}
    \Photon(0,5)(30,5){2}{6.5}
  \end{picture}}}
}
\newcommand{\Gluonline}[1]{
\vcenter{\hbox{
  \begin{picture}(60,20)(0,{#1})
  \SetScale{2.}
    \Gluon(0,5)(30,5){2}{6.5}
  \end{picture}}}
}
\newcommand{\Wbosline}[1]{
\vcenter{\hbox{
  \begin{picture}(60,20)(0,{#1})
  \SetScale{2.}
    \Photon(0,5)(30,5){2}{4}
    \ArrowLine(13.3,3.1)(16.9,7.2)
  \end{picture}}}
}
\newcommand{\Zbosline}[1]{
\vcenter{\hbox{
  \begin{picture}(60,20)(0,{#1})
  \SetScale{2.}
    \Photon(0,5)(30,5){2}{4}
  \end{picture}}}
}
\newcommand{\Philine}[1]{
\vcenter{\hbox{
  \begin{picture}(60,20)(0,{#1})
  \SetScale{2.}
    \DashLine(0,5)(30,5){2}
  \end{picture}}}
}
\newcommand{\Phicline}[1]{
\vcenter{\hbox{
  \begin{picture}(60,20)(0,{#1})
  \SetScale{2.}
    \DashLine(0,5)(30,5){2}
    \ArrowLine(14,5)(16,5)
  \end{picture}}}
}
\newcommand{\Ghostline}[1]{
\vcenter{\hbox{
  \begin{picture}(60,20)(0,{#1})
  \SetScale{2.}
    \DashLine(0,5)(30,5){.5}
    \ArrowLine(14,5)(16,5)
  \end{picture}}}
}
%
% SM Lagrangian in \Rxi
%======================
%
\newcommand{\gauge}{g}
\newcommand{\gpar}{\xi}
\newcommand{\gparA}{\xi_{_A}}
\newcommand{\gparZ}{\xi_{_Z}}
\newcommand{\gpari}[1]{\gpar_{#1}}
\newcommand{\gparis}[1]{\gpar^2_{#1}}
\newcommand{\gpariq}[1]{\gpar^4_{#1}}
\newcommand{\gpars}{\xi^2}
\newcommand{\dgpar}{\Delta\gpar}
\newcommand{\dgparA}{\Delta\gparA}
\newcommand{\dgparZ}{\Delta\gparZ}
\newcommand{\gparq}{\xi^4}
\newcommand{\gparAs}{\xi^2_{_A}}
\newcommand{\gparAq}{\xi^4_{_A}}
\newcommand{\gparZs}{\xi^2_{_Z}}
\newcommand{\gparZq}{\xi^4_{_Z}}
\newcommand{\Rxi}{R_{\gpar}}
\newcommand{\UG}{U}
\newcommand{\UGi}{\ssU}
\newcommand{\hxi}{\chi}
%
% Lagrangiains
%-------------
%
\newcommand{\LSM}{{\cal{L}}_{_{\rm{SM}}}}
\newcommand{\LSMr}{{\cal{L}}^{\rm{\ssR}}_{_{\rm{SM}}}}
\newcommand{\LYM}{{\cal{L}}_{_{\rm{YM}}}}
\newcommand{\Lzer}{{\cal{L}}_{0}}
\newcommand{\Lone}{{\cal{L}}^{{\bos},{\rm{I}}}}
\newcommand{\Lpro}{{\cal{L}}_{\rm{prop}}}
\newcommand{\Ls  }{{\cal{L}}_{_{\rm{S}}}}
\newcommand{\Lsi }{{\cal{L}}^{\rm{I}}_{_{\rm{S}}}}
\newcommand{\Lgf }{{\cal{L}}_{\rm{gf}}}
\newcommand{\Lgfi}{{\cal{L}}^{\rm{I}}_{\rm{gf}}}
\newcommand{\Lf  }{{\cal{L}}^{{\fer},{\rm{I}}}_{\ssV}}
\newcommand{\LHf }{{\cal{L}}^{\fer}_{\ssS}}
\newcommand{\LHfm}{{\cal{L}}^{{\fer},m}_{\ssS}}
\newcommand{\LHfi}{{\cal{L}}^{{\fer},{\rm{I}}}_{\ssS}}
\newcommand{\Lren}{{\cal{L}}_{\rm{\ssR}}}
\newcommand{\Lct}{{\cal{L}}_{\rm{ct}}}
\newcommand{\Lcti}[1]{{\cal{L}}^{#1}_{\rm{ct}}}
\newcommand{\LctI}{{\cal{L}}^{(2)}_{\rm{ct}}}
\newcommand{\Llone}{{\cal{L}}}
\newcommand{\LQED}{{\cal{L}}_{_{\rm{QED}}}}
\newcommand{\LQEDz}{{\cal{L}}^{0}_{_{\rm{QED}}}}
\newcommand{\LQEDi}{{\cal{L}}^{\rm{\ssI}}_{_{\rm{QED}}}}
\newcommand{\LQEDr}{{\cal{L}}^{\rm{\ssR}}_{_{\rm{QED}}}}
\newcommand{\Greenf}{G}
\newcommand{\Greenfa}[1]{G\lpar{#1}\rpar}
\newcommand{\Greenft}[2]{G\lpar{#1,#2}\rpar}
\newcommand{\FST}[3]{F_{#1#2}^{#3}}
\newcommand{\cD}[1]{D_{#1}}
\newcommand{\pd}[1]{\partial_{#1}}
\newcommand{\tgen}[1]{\tau^{#1}}
\newcommand{\gbl}{g_1}
\newcommand{\lctt}[3]{\varepsilon_{#1#2#3}}
\newcommand{\lctf}[4]{\varepsilon_{#1#2#3#4}}
\newcommand{\lctfb}[4]{\varepsilon\lpar{#1#2#3#4}\rpar}
\newcommand{\slct}{\varepsilon}
%gauge fixing
\newcommand{\cgfi}[1]{{\cal{C}}^{#1}}
\newcommand{\cgfZ}{{\cal{C}}^{\ssZ}}
\newcommand{\cgfA}{{\cal{C}}^{\ssA}}
% Temporary change !!!!!!!!!!
%\newcommand{\cgfZs}{{\cal{C}}^2_{_Z}}
%\newcommand{\cgfAs}{{\cal{C}}^2_{_A}}
%parameters of scalar potential
\newcommand{\hpms}{\mu^2}
\newcommand{\hpal}{\alpha_{_H}}
\newcommand{\hpals}{\alpha^2_{_H}}
\newcommand{\hpbe}{\beta_{_H}}
\newcommand{\hpbep}{\beta^{'}_{_H}}
\newcommand{\hpla}{\lambda}
\newcommand{\hpalf}{\alpha_{f}}
\newcommand{\hpbef}{\beta_{f}}
%transformation parameters
\newcommand{\tpar}[1]{\Lambda^{#1}}
%M,L-operators
\newcommand{\Mop}[2]{{\rm{M}}^{#1#2}}
\newcommand{\Lop}[2]{{\rm{L}}^{#1#2}}
\newcommand{\Lgen}[1]{T^{#1}}
\newcommand{\Rgen}[1]{t^{#1}}
\newcommand{\fpari}[1]{\lambda_{#1}}
\newcommand{\fQ}[1]{Q_{#1}}
\newcommand{\unm}{I}
\newcommand{\cDsla}{/\!\!\!\!D}
%
% A-B-C-D functions
%==================
%
\newcommand{\saff}[1]{A_{#1}}                    % A form-factors
\newcommand{\aff}[2]{A_{#1}\lpar #2\rpar}                   
\newcommand{\sbff}[1]{B_{#1}}                    % B form-factors
\newcommand{\sfbff}[1]{B^{\ssF}_{#1}}
\newcommand{\bff}[4]{B_{#1}\lpar #2;#3,#4\rpar}             
\newcommand{\bfft}[3]{B_{#1}\lpar #2,#3\rpar}             
\newcommand{\fbff}[4]{B^{\ssF}_{#1}\lpar #2;#3,#4\rpar}        
\newcommand{\cdbff}[4]{\Delta B_{#1}\lpar #2;#3,#4\rpar}             
\newcommand{\sdbff}[4]{\delta B_{#1}\lpar #2;#3,#4\rpar}             
\newcommand{\cdbfft}[3]{\Delta B_{#1}\lpar #2,#3\rpar}             
\newcommand{\sdbfft}[3]{\delta B_{#1}\lpar #2,#3\rpar}             
\newcommand{\scff}[1]{C_{#1}}                    % C form-factors
\newcommand{\scffo}[2]{C_{#1}\lpar{#2}\rpar}                
\newcommand{\cff}[7]{C_{#1}\lpar #2,#3,#4;#5,#6,#7\rpar}    
\newcommand{\sccff}[5]{c_{#1}\lpar #2;#3,#4,#5\rpar} 
\newcommand{\sdff}[1]{D_{#1}}                    % D form-factors
\newcommand{\dffp}[7]{D_{#1}\lpar #2,#3,#4,#5,#6,#7;}       
\newcommand{\dffm}[4]{#1,#2,#3,#4\rpar}                     
\newcommand{\bzfa}[2]{B^{\ssF}_{_{#2}}\lpar{#1}\rpar}
\newcommand{\bzfaa}[3]{B^{\ssF}_{_{#2#3}}\lpar{#1}\rpar}
\newcommand{\shcff}[4]{C_{_{#2#3#4}}\lpar{#1}\rpar}
\newcommand{\shdff}[6]{D_{_{#3#4#5#6}}\lpar{#1,#2}\rpar}
\newcommand{\scdff}[3]{d_{#1}\lpar #2,#3\rpar} 
\newcommand{\scaldff}[1]{{\cal{D}}^{#1}}
\newcommand{\caldff}[2]{{\cal{D}}^{#1}\lpar{#2}\rpar}
\newcommand{\caldfft}[3]{{\cal{D}}_{#1}^{#2}\lpar{#3}\rpar}
%
% a-b-c-d functions
%------------------
%
\newcommand{\slaff}[1]{a_{#1}}                        
\newcommand{\slbff}[1]{b_{#1}}                        
\newcommand{\slbffh}[1]{{\hat{b}}_{#1}}    
\newcommand{\ssldff}[1]{d_{#1}}                        
\newcommand{\sslcff}[1]{c_{#1}}                        
\newcommand{\slcff}[2]{c_{#1}^{(#2)}}                        
\newcommand{\sldff}[2]{d_{#1}^{(#2)}}                        
\newcommand{\lbff}[3]{b_{#1}\lpar #2;#3\rpar}         
\newcommand{\lbffh}[2]{{\hat{b}}_{#1}\lpar #2\rpar}   
\newcommand{\lcff}[8]{c_{#1}^{(#2)}\lpar  #3,#4,#5;#6,#7,#8\rpar}         
\newcommand{\ldffp}[8]{d_{#1}^{(#2)}\lpar #3,#4,#5,#6,#7,#8;}
\newcommand{\ldffm}[4]{#1,#2,#3,#4\rpar}                   
%
% I-J functions
%--------------
%
\newcommand{\Iff}[4]{I_{#1}\lpar #2;#3,#4 \rpar}
\newcommand{\Jff}[4]{J_{#1}\lpar #2;#3,#4 \rpar}
\newcommand{\Jds}[5]{{\bar{J}}_{#1}\lpar #2,#3;#4,#5 \rpar}
\newcommand{\sJds}[1]{{\bar{J}}_{#1}}
%--
% n-dimension and epsilons
%=========================
%
\newcommand{\nhmt}{\frac{n}{2}-2}
\newcommand{\nhmts}{{n}/{2}-2}
\newcommand{\omnh}{1-\frac{n}{2}}
\newcommand{\nhmo}{\frac{n}{2}-1}
\newcommand{\fmon}{4-n}
\newcommand{\lpi}{\ln\pi}
\newcommand{\lmass}[1]{\ln #1}
\newcommand{\egnh}{\egam{\frac{n}{2}}}
\newcommand{\egomnh}{\egam{1-\frac{n}{2}}}
\newcommand{\egtmnh}{\egam{2-\frac{n}{2}}}
\newcommand{\Ddr}{{\ds\frac{1}{{\bar{\varepsilon}}}}}
\newcommand{\Ddrs}{{\ds\frac{1}{{\bar{\varepsilon}^2}}}}
\newcommand{\Ddrd}{{\bar{\varepsilon}}}
\newcommand{\ept}{\hat\varepsilon}
\newcommand{\Ddrh}{{\ds\frac{1}{\hat{\varepsilon}}}}
\newcommand{\Ddrp}{{\ds\frac{1}{\varepsilon'}}}
\newcommand{\Ddrps}{\lpar{\ds{\frac{1}{\varepsilon'}}}\rpar^2}
\newcommand{\dre}{\varepsilon}
\newcommand{\drei}[1]{\varepsilon_{#1}}
\newcommand{\epp}{\varepsilon'}
\newcommand{\eps}{\varepsilon^*}
%--
%-- Im for masses and propagators
%================================
\newcommand{\ep}{\epsilon}
%--
\newcommand{\propbt}[6]{{{#1_{#2}#1_{#3}}\over{\lpar #1^2 + #4 
-\ib\ep\rpar\lpar\lpar #5\rpar^2 + #6 -\ib\ep\rpar}}}
\newcommand{\propbo}[5]{{{#1_{#2}}\over{\lpar #1^2 + #3 - \ib\ep\rpar
\lpar\lpar #4\rpar^2 + #5 -\ib\ep\rpar}}}
\newcommand{\propc}[6]{{1\over{\lpar #1^2 + #2 - \ib\ep\rpar
\lpar\lpar #3\rpar^2 + #4 -\ib\ep\rpar
\lpar\lpar #5\rpar^2 + #6 -\ib\ep\rpar}}}
%--
\newcommand{\propa}[2]{{1\over {#1^2 + #2^2 - \ib\ep}}}
\newcommand{\propb}[4]{{1\over {\lpar #1^2 + #2 - \ib\ep\rpar
\lpar\lpar #3\rpar^2 + #4 -\ib\ep\rpar}}}
\newcommand{\propbs}[4]{{1\over {\lpar\lpar #1\rpar^2 + #2 - \ib\ep\rpar
\lpar\lpar #3\rpar^2 + #4 -\ib\ep\rpar}}}
\newcommand{\propat}[4]{{#3_{#1}#3_{#2}\over {#3^2 + #4^2 - \ib\ep}}}
\newcommand{\propaf}[6]{{#5_{#1}#5_{#2}#5_{#3}#5_{#4}\over 
{#5^2 + #6^2 -\ib\ep}}}
\newcommand{\momeps}[1]{#1^2 - \ib\ep}
\newcommand{\mopeps}[1]{#1^2 + \ib\ep}
%--
\newcommand{\propz}[1]{{1\over{#1^2 + \mzs - \ib\ep}}}
\newcommand{\propw}[1]{{1\over{#1^2 + \mws - \ib\ep}}}
\newcommand{\proph}[1]{{1\over{#1^2 + \mhs - \ib\ep}}}
\newcommand{\propf}[2]{{1\over{#1^2 + #2}}}
\newcommand{\propzrg}[3]{{{\delta_{#1#2}}\over{{#3}^2 + \mzs - \ib\ep}}}
\newcommand{\propwrg}[3]{{{\delta_{#1#2}}\over{{#3}^2 + \mws - \ib\ep}}}
\newcommand{\propzug}[3]{{
      {\delta_{#1#2} + \displaystyle{{{#3}^{#1}{#3}^{#2}}\over{\mzs}}}
                         \over{{#3}^2 + \mzs - \ib\ep}}}
\newcommand{\propwug}[3]{{
      {\delta_{#1#2} + \displaystyle{{{#3}^{#1}{#3}^{#2}}\over{\mws}}}
                        \over{{#3}^2 + \mws - \ib\ep}}}
%----------------------------------------
\newcommand{\thf}[1]{\theta\lpar #1\rpar}
\newcommand{\epf}[1]{\varepsilon\lpar #1\rpar}
\newcommand{\singp}{\stackrel{\rm{sing}}{\rightarrow}}
\newcommand{\aint}[3]{\int_{#1}^{#2}\,d #3}
\newcommand{\aroot}[1]{\sqrt{#1}}
\newcommand{\gramc}{\Delta_3}
\newcommand{\gramd}{\Delta_4}
\newcommand{\pinch}[2]{P^{(#1)}\lpar #2\rpar}
\newcommand{\pinchc}[2]{C^{(#1)}_{#2}}
\newcommand{\pinchd}[2]{D^{(#1)}_{#2}}
\newcommand{\loarg}[1]{\ln\lpar #1\rpar}
\newcommand{\loargr}[1]{\ln\lrbr #1\rrbr}
\newcommand{\lsoarg}[1]{\ln^2\lpar #1\rpar}
\newcommand{\ltarg}[2]{\ln\lpar #1\rpar\lpar #2\rpar}
\newcommand{\rfun}[2]{R\lpar #1,#2\rpar}
\newcommand{\pinchb}[3]{B_{#1}\lpar #2,#3\rpar}
\newcommand{\lga}{\ph}
\newcommand{\lzga}{\ssZ\ph}
%
% Auxiliary functions
%
\newcommand{\afa}[5]{A_{#1}^{#2}\lpar #3;#4,#5\rpar}
\newcommand{\bfa}[5]{B_{#1}^{#2}\lpar #3;#4,#5\rpar} 
\newcommand{\hfa}[5]{H_{#1}^{#2}\lpar #3;#4,#5\rpar}
\newcommand{\rfa}[5]{R_{#1}^{#2}\lpar #3;#4,#5\rpar}
\newcommand{\afao}[3]{A_{#1}^{#2}\lpar #3\rpar}
\newcommand{\bfao}[3]{B_{#1}^{#2}\lpar #3\rpar}
\newcommand{\hfao}[3]{H_{#1}^{#2}\lpar #3\rpar}
\newcommand{\rfao}[3]{R_{#1}^{#2}\lpar #3\rpar}
\newcommand{\afax}[6]{A_{#1}^{#2}\lpar #3;#4,#5,#6\rpar}
\newcommand{\afas}[2]{A_{#1}^{#2}}
\newcommand{\bfas}[2]{B_{#1}^{#2}}
\newcommand{\hfas}[2]{H_{#1}^{#2}}
\newcommand{\rfas}[2]{R_{#1}^{#2}}
\newcommand{\tfas}[2]{T_{#1}^{#2}}
\newcommand{\afaR}[6]{A_{#1}^{\gpar}\lpar #2;#3,#4,#5,#6 \rpar}
\newcommand{\bfaR}[6]{B_{#1}^{\gpar}\lpar #2;#3,#4,#5,#6 \rpar}
\newcommand{\hfaR}[6]{H_{#1}^{\gpar}\lpar #2;#3,#4,#5,#6 \rpar}
\newcommand{\shfaR}[1]{H_{#1}^{\gpar}}
\newcommand{\rfaR}[6]{R_{#1}^{\gpar}\lpar #2;#3,#4,#5,#6 \rpar}
\newcommand{\srfaR}[1]{R_{#1}^{\gpar}}
\newcommand{\afaRg}[5]{A_{#1 \lga}^{\gpar}\lpar #2;#3,#4,#5 \rpar}
\newcommand{\bfaRg}[5]{B_{#1 \lga}^{\gpar}\lpar #2;#3,#4,#5 \rpar}
\newcommand{\hfaRg}[5]{H_{#1 \lga}^{\gpar}\lpar #2;#3,#4,#5 \rpar}
\newcommand{\shfaRg}[1]{H_{#1\lga}^{\gpar}}
\newcommand{\rfaRg}[5]{R_{#1 \lga}^{\gpar}\lpar #2;#3,#4,#5 \rpar}
\newcommand{\srfaRg}[1]{R_{#1\lga}^{\gpar}}
\newcommand{\afaRt}[3]{A_{#1}^{\gpar}\lpar #2,#3 \rpar}
\newcommand{\hfaRt}[3]{H_{#1}^{\gpar}\lpar #2,#3 \rpar}
\newcommand{\hfaRf}[4]{H_{#1}^{\gpar}\lpar #2,#3,#4 \rpar}
\newcommand{\afasm}[4]{A_{#1}^{\lpar #2,#3,#4 \rpar}}
\newcommand{\bfasm}[4]{B_{#1}^{\lpar #2,#3,#4 \rpar}}
\newcommand{\color}[1]{c_{#1}}
\newcommand{\htf}[2]{H_2\lpar #1,#2\rpar}
\newcommand{\rof}[2]{R_1\lpar #1,#2\rpar}
\newcommand{\rtf}[2]{R_3\lpar #1,#2\rpar}
\newcommand{\rtrans}[2]{R_{#1}^{#2}}
\newcommand{\momf}[2]{#1^2_{#2}}
\newcommand{\Scalvert}[8][70]{
  \vcenter{\hbox{
  \SetScale{0.8}
  \begin{picture}(#1,50)(15,15)
    \Line(25,25)(50,50)      \Text(34,20)[lc]{#6} \Text(11,20)[lc]{#3}
    \Line(50,50)(25,75)      \Text(34,60)[lc]{#7} \Text(11,60)[lc]{#4}
    \Line(50,50)(90,50)      \Text(11,40)[lc]{#2} \Text(55,33)[lc]{#8}
    \GCirc(50,50){10}{1}          \Text(60,48)[lc]{#5} 
  \end{picture}}}
  }
%
% db-s additions, beware, I've modified above also
%
\newcommand{\tHs}{\mu}
\newcommand{\tHsz}{\mu_{_0}}
\newcommand{\tHss}{\mu^2}
\newcommand{\Reb}{{\rm{Re}}}
\newcommand{\Imb}{{\rm{Im}}}
%
% gp's additions 
%
\newcommand{\spd}{\partial}
\newcommand{\fun}[1]{f\lpar{#1}\rpar}
\newcommand{\ffun}[2]{F_{#1}\lpar #2\rpar}
\newcommand{\gfun}[2]{G_{#1}\lpar #2\rpar}
\newcommand{\sffun}[1]{F_{#1}}
\newcommand{\csffun}[1]{{\cal{F}}_{#1}}
\newcommand{\sgfun}[1]{G_{#1}}
\newcommand{\tpfi}{\lpar 2\pi\rpar^4\ib}
\newcommand{\ffv}{F_{_V}}
\newcommand{\fga}{G_{_A}}
\newcommand{\ffm}{F_{_M}}
\newcommand{\ffs}{F_{_S}}
\newcommand{\fgp}{G_{_P}}
\newcommand{\fge}{G_{_E}}
\newcommand{\ffa}{F_{_A}}
\newcommand{\ffps}{F_{_P}}
\newcommand{\ffe}{F_{_E}}
\newcommand{\gacom}[2]{\lpar #1 + #2\gfd\rpar}
\newcommand{\mft}{m_{\tilde f}}
\newcommand{\qft}{Q_{f'}}
\newcommand{\vft}{v_{\tilde f}}
\newcommand{\subb}[2]{b_{#1}\lpar #2 \rpar}
\newcommand{\fwfr}[5]{\Sigma\lpar #1,#2,#3;#4,#5 \rpar}
\newcommand{\slim}[2]{\lim_{#1 \to #2}}
\newcommand{\sprop}[3]{
{#1\over {\lpar q^2\rpar^2\lpar \lpar q+ #2\rpar^2+#3^2\rpar }}}
%
% roots, variables, coefficients
%
\newcommand{\xroot}[1]{x_{#1}}
\newcommand{\yroot}[1]{y_{#1}}
\newcommand{\zroot}[1]{z_{#1}}
\newcommand{\lvar}{l}
\newcommand{\rvar}{r}
\newcommand{\tvar}{t}
\newcommand{\uvar}{u}
\newcommand{\vvar}{v}
\newcommand{\xvar}{x}
\newcommand{\yvar}{y}
\newcommand{\zvar}{z}
\newcommand{\xvarp}{x'}
\newcommand{\yvarp}{y'}
\newcommand{\zvarp}{z'}
\newcommand{\rvars}{r^2}
\newcommand{\vvars}{v^2}
\newcommand{\xvars}{x^2}
\newcommand{\yvars}{y^2}
\newcommand{\zvars}{z^2}
\newcommand{\rvarc}{r^3}
\newcommand{\xvarc}{x^3}
\newcommand{\yvarc}{y^3}
\newcommand{\zvarc}{z^3}
\newcommand{\rvarq}{r^4}
\newcommand{\xvarq}{x^4}
\newcommand{\yvarq}{y^4}
\newcommand{\zvarq}{z^4}
\newcommand{\avar}{a}
\newcommand{\avars}{a^2}
\newcommand{\avarc}{a^3}
\newcommand{\avari}[1]{a_{#1}}
\newcommand{\avart}[2]{a_{#1}^{#2}}
\newcommand{\delvari}[1]{\delta_{#1}}
\newcommand{\rvari}[1]{r_{#1}}
\newcommand{\xvari}[1]{x_{#1}}
\newcommand{\yvari}[1]{y_{#1}}
\newcommand{\zvari}[1]{z_{#1}}
\newcommand{\rvart}[2]{r_{#1}^{#2}}
\newcommand{\xvart}[2]{x_{#1}^{#2}}
\newcommand{\yvart}[2]{y_{#1}^{#2}}
\newcommand{\zvart}[2]{z_{#1}^{#2}}
\newcommand{\rvaris}[1]{r^2_{#1}}
\newcommand{\xvaris}[1]{x^2_{#1}}
\newcommand{\yvaris}[1]{y^2_{#1}}
\newcommand{\zvaris}[1]{z^2_{#1}}
\newcommand{\Xvar}{X}
\newcommand{\Xvars}{X^2}
\newcommand{\Xvari}[1]{X_{#1}}
\newcommand{\Xvaris}[1]{X^2_{#1}}
\newcommand{\Yvar}{Y}
\newcommand{\Yvars}{Y^2}
\newcommand{\Yvari}[1]{Y_{#1}}
\newcommand{\Yvaris}[1]{Y^2_{#1}}
%---
\newcommand{\lnx}{\ln\xvar}
\newcommand{\lnz}{\ln\zvar}
\newcommand{\lnsx}{\ln^2\xvar}
\newcommand{\lnsz}{\ln^2\zvar}
\newcommand{\lncz}{\ln^3\zvar}
\newcommand{\lnomz}{\ln\lpar 1-\zvar\rpar}
\newcommand{\lnsomz}{\ln^2\lpar 1-\zvar\rpar}
\newcommand{\ccoefi}[1]{c_{#1}}
\newcommand{\ccoeft}[2]{c^{#1}_{#2}}
%
% Matrices
%
\newcommand{\Smat}{{\cal{S}}}
\newcommand{\Mmat}{{\cal{M}}}
\newcommand{\Rmat}{{\cal{R}}}
\newcommand{\Xmat}[1]{X_{#1}}
\newcommand{\XmatI}[1]{X^{-1}_{#1}}
\newcommand{\unitmat}{I}
\newcommand{\zeromat}{O}
\newcommand{\paulimat}[1]{\tau_{#1}}
\newcommand{\Umat  }{U}
\newcommand{\Umath }{U^{+}}
\newcommand{\UmatL }{{\cal{U}}_{\ssL}}
\newcommand{\UmatLh}{{\cal{U}}^{+}_{\ssL}}
\newcommand{\UmatR }{{\cal{U}}_{\ssR}}
\newcommand{\UmatRh}{{\cal{U}}^{+}_{\ssR}}
\newcommand{\Vmat  }{V}
\newcommand{\Vmath }{V^{+}}
\newcommand{\VmatL }{{\cal{D}}_{\ssL}}
\newcommand{\VmatLh}{{\cal{D}}^{+}_{\ssL}}
\newcommand{\VmatR }{{\cal{D}}_{\ssR}}
\newcommand{\VmatRh}{{\cal{D}}^{+}_{\ssR}}
\newcommand{\Kmat}{{C}}
\newcommand{\Kmatc}{{C}^{\dagger}}
\newcommand{\Kmati}[1]{{C}_{#1}}
\newcommand{\Kmatci}[1]{{C}^{\dagger}_{#1}}
\newcommand{\ffac}[2]{f_{#1}^{#2}}
\newcommand{\Ffac}[1]{F_{#1}}
\newcommand{\Rvec}[2]{R^{(#1)}_{#2}}
\newcommand{\momfl}[2]{#1_{#2}}
\newcommand{\momfs}[2]{#1^2_{#2}}
\newcommand{\fpseZ}{A^{\ssF\ssP,\ssZ}}
\newcommand{\fpseA}{A^{\ssF\ssP,\ssA}}
\newcommand{\fptZ}{T^{\ssF\ssP,\ssZ}}
\newcommand{\fptA}{T^{\ssF\ssP,\ssA}}
\newcommand{\dprop}{\overline\Delta}
\newcommand{\dpropi}[1]{d_{#1}}
\newcommand{\dpropic}[1]{d^{c}_{#1}}
\newcommand{\dpropii}[2]{d_{#1}\lpar #2\rpar}
\newcommand{\dpropis}[1]{d^2_{#1}}
\newcommand{\dproppi}[1]{d'_{#1}}
\newcommand{\psf}[4]{P\lpar #1;#2,#3,#4\rpar}
\newcommand{\ssf}[5]{S^{(#1)}\lpar #2;#3,#4,#5\rpar}
\newcommand{\csf}[5]{C_{_S}^{(#1)}\lpar #2;#3,#4,#5\rpar}
%
% polarization vectors
%=====================
%
\newcommand{\lvec}{l}
\newcommand{\lvecs}{l^2}
\newcommand{\lveci}[1]{l_{#1}}
\newcommand{\mvec}{m}
\newcommand{\mvecs}{m^2}
\newcommand{\mveci}[1]{m_{#1}}
\newcommand{\nvec}{n}
\newcommand{\nvecs}{n^2}
\newcommand{\nveci}[1]{n_{#1}}
\newcommand{\epi}[1]{\epsilon_{#1}}
\newcommand{\phep}[1]{\ep_{#1}}
\newcommand{\php}[3]{\ep^{#1}_{#2}\lpar #3 \rpar}
\newcommand{\sphep}{\ep}
\newcommand{\vbep}[1]{e_{#1}}
\newcommand{\vbepp}[1]{e^{+}_{#1}}
\newcommand{\vbepm}[1]{e^{-}_{#1}}
\newcommand{\svbep}{e}
%
% longitudinal polarizations
%===========================
%
\newcommand{\lpol}{\lambda}
\newcommand{\spol}{\sigma}
\newcommand{\rpol}{\rho  }
\newcommand{\kpol}{\kappa}
\newcommand{\lpols}{\lambda^2}
\newcommand{\spols}{\sigma^2}
\newcommand{\rpols}{\rho^2}
\newcommand{\kpols}{\kappa^2}
\newcommand{\lpoli}[1]{\lambda_{#1}}
\newcommand{\spoli}[1]{\sigma_{#1}}
\newcommand{\rpoli}[1]{\rho_{#1}}
\newcommand{\kpoli}[1]{\kappa_{#1}}
%
% some vectors
%=============
%
\newcommand{\uvec}{u}
\newcommand{\uveci}[1]{u_{#1}}
%
% Momenta:
%=========
%
\newcommand{\imom}{q}
\newcommand{\imomi}[1]{q_{#1}}
\newcommand{\imoms}{q^2}
\newcommand{\pmom}{p}
\newcommand{\pmomp}{p'}
\newcommand{\pmoms}{p^2}
\newcommand{\pmomq}{p^4}
\newcommand{\pmomx}{p^6}
\newcommand{\pmomi}[1]{p_{#1}}
\newcommand{\pmompi}[1]{p'_{#1}}
\newcommand{\pmomis}[1]{p^2_{#1}}
%--
\newcommand{\Pmom}{P}
\newcommand{\Pmoms}{P^2}
\newcommand{\Pmomi}[1]{P_{#1}}
\newcommand{\Pmomis}[1]{P^2_{#1}}
%--
\newcommand{\Pmomp}{P'}
\newcommand{\Pmomps}{{P'}^2}
\newcommand{\Pmompi}[1]{P'_{#1}}
\newcommand{\Pmompis}[1]{{P'}^2_{#1}}
\newcommand{\Kmom}{K}
\newcommand{\Kmoms}{K^2}
\newcommand{\Kmomi}[1]{K_{#1}}
\newcommand{\Kmomis}[1]{K^2_{#1}}
%--
\newcommand{\kmom}{k}
\newcommand{\kmoms}{k^2}
\newcommand{\kmomi}[1]{k_{#1}}
\newcommand{\lmom}{l}
\newcommand{\lmoms}{l^2}
\newcommand{\lmomi}[1]{l_{#1}}
\newcommand{\qmom}{q}
\newcommand{\qmoms}{q^2}
\newcommand{\qmomi}[1]{q_{#1}}
\newcommand{\qmomis}[1]{q^2_{#1}}
\newcommand{\smom}{s}
\newcommand{\smoms}{s^2}
\newcommand{\smomi}[1]{s_{#1}}
\newcommand{\tmom}{t}
\newcommand{\tmoms}{t^2}
\newcommand{\tmomi}[1]{t_{#1}}
\newcommand{\Trmom}{Q}
\newcommand{\Prmom}{P}
\newcommand{\gmv}{Q^2}
\newcommand{\Trmoms}{Q^2}
\newcommand{\Prmoms}{P^2}
\newcommand{\Ptmoms}{T^2}
\newcommand{\Pumoms}{U^2}
\newcommand{\Trmomq}{Q^4}
\newcommand{\Prmomq}{P^4}
\newcommand{\Ptmomq}{T^4}
\newcommand{\Pumomq}{U^4}
\newcommand{\Trmomx}{Q^6}
\newcommand{\Trmomi}[1]{Q_{#1}}
\newcommand{\Trmomis}[1]{Q^2_{#1}}
\newcommand{\Prmomi}[1]{P_{#1}}
\newcommand{\pone}{p_1}
\newcommand{\ptwo}{p_2}
\newcommand{\ptre}{p_3}
\newcommand{\pfor}{p_4}
\newcommand{\pones}{p_1^2}
\newcommand{\ptwos}{p_2^2}
\newcommand{\ptres}{p_3^2}
\newcommand{\pfors}{p_4^2}
\newcommand{\poneq}{p_1^4}
\newcommand{\ptwoq}{p_2^4}
\newcommand{\ptreq}{p_3^4}
\newcommand{\pforq}{p_4^4}
\newcommand{\modmom}[1]{\mid{\vec{#1}}\mid}
\newcommand{\modmoms}[1]{\mid{\vec{#1}}\mid^2}
\newcommand{\modmomi}[2]{\mid{\vec{#1}}_{#2}\mid}
\newcommand{\vect}[1]{{\vec{#1}}}
\newcommand{\Energ}{E}
\newcommand{\Energp}{E'}
\newcommand{\Energpp}{E''}
\newcommand{\Energs}{E^2}
\newcommand{\Energc}{E^3}
\newcommand{\Energf}{E^4}
\newcommand{\Energv}{E^5}
\newcommand{\Energx}{E^6}
\newcommand{\Energi}[1]{E_{#1}}
\newcommand{\Energt}[2]{E_{#1}^{#2}}
\newcommand{\Energis}[1]{E^2_{#1}}
\newcommand{\energ}{e}
\newcommand{\energp}{e'}
\newcommand{\energpp}{e''}
\newcommand{\energs}{e^2}
\newcommand{\energi}[1]{e_{#1}}
\newcommand{\energt}[2]{e_{#1}^{#2}}
\newcommand{\energis}[1]{e^2_{#1}}
\newcommand{\wenerg}{w}
\newcommand{\wenergs}{w^2}
\newcommand{\wenergi}[1]{w_{#1}}
\newcommand{\wenergp}{w'}
\newcommand{\wenergpp}{w''}
%
% kinematical cuts
%=================
%
\newcommand{\ecut}{e}
\newcommand{\ecuts}{e^2}
\newcommand{\ecuti}[1]{e^{#1}}
\newcommand{\ccut}{c_m}
\newcommand{\ccuti}[1]{c_{#1}}
\newcommand{\ccuts}{c^2_m}
\newcommand{\scuts}{s^2_m}
\newcommand{\ccutis}[1]{c^2_{#1}}
\newcommand{\ccutic}[1]{c^3_{#1}}
\newcommand{\ccutc}{c^3_m}
\newcommand{\rcut}{\varrho}
\newcommand{\rcuts}{\varrho^2}
\newcommand{\rcuti}[1]{\varrho_{#1}}
\newcommand{\rcutu}[1]{\varrho^{#1}}
\newcommand{\Dcut}{\Delta}
%
%-----LIB_VERT_XI1.TEX----------------
\newcommand{\dwf}{\delta_{_{\rm{WF}}}}
\newcommand{\gbar}{\overline g}
\newcommand{\PP}{\mbox{PP}}
\newcommand{\mv}{m_{_V}}
\newcommand{\bGv}{{\overline\Gamma}_{_V}}
\newcommand{\Umuv}{\hat{\mu}_\ssV}
\newcommand{\Svv}{{\Sigma}_\ssV}
\newcommand{\muv}{p_\ssV}
\newcommand{\muvb}{\mu_{\ssV_{0}}}
\newcommand{\URPvv}{{P}_\ssV}
\newcommand{\RPvv}{{P}_\ssV}
\newcommand{\Svvrem}{{\Sigma}_\ssV^{\mathrm{rem}}}
\newcommand{\USvvrem}{\hat{\Sigma}_\ssV^{\mathrm{rem}}}
\newcommand{\Gv}{\Gamma_{_V}}
%
% Renormalization 
%
\newcommand{\param}{p}
\newcommand{\parami}[1]{p^{#1}}
\newcommand{\paramb}{p_{0}}
\newcommand{\Zcon}{Z}
\newcommand{\Zconi}[1]{Z_{#1}}
\newcommand{\zconi}[1]{z_{#1}}
\newcommand{\Zconim}[1]{{Z^-_{#1}}}
\newcommand{\zconim}[1]{{z^-_{#1}}}
\newcommand{\Zcont}[2]{Z_{#1}^{#2}}
\newcommand{\zcont}[2]{z_{#1}^{#2}}
\newcommand{\zcontm}[2]{z_{#1}^{{#2}-}}
\newcommand{\sZconi}[2]{\sqrt{Z_{#1}}^{\;#2}}
\newcommand{\gacome}[1]{\lpar #1 - \gfd\rpar}
\newcommand{\sPj}[2]{\Lambda^{#1}_{#2}}
\newcommand{\sPjs}[2]{\Lambda_{#1,#2}}
\newcommand{\amos}{\mbox{$M^2_{_1}$}}
\newcommand{\amts}{\mbox{$M^2_{_2}$}}
\newcommand{\er}{e_{_{R}}}
\newcommand{\epr}{e'_{_{R}}}
\newcommand{\ers}{e^2_{_{R}}}
\newcommand{\erc}{e^3_{_{R}}}
\newcommand{\erq}{e^4_{_{R}}}
\newcommand{\erf}{e^5_{_{R}}}
\newcommand{\sour}{J}
\newcommand{\sourb}{\overline J}
\newcommand{\lrm}{M_{_R}}
%
% db: fermionic self-energies and vertex libraries
%
\newcommand{\vlami}[1]{\lambda_{#1}}
\newcommand{\vlamis}[1]{\lambda^2_{#1}}
\newcommand{\Vvert}{V}
\newcommand{\Avert}{A}
\newcommand{\Svert}{S}
\newcommand{\Pvert}{P}
\newcommand{\vvert}{F}
\newcommand{\Cvert}{\cal{V}}
\newcommand{\Bvert}{\cal{B}}
\newcommand{\Vveri}[2]{V_{#1}^{#2}}
\newcommand{\Fveri}[1]{{\cal{F}}^{#1}}
\newcommand{\Cveri}[1]{{\cal{V}}\lpar{#1}\rpar}
\newcommand{\Bveri}[1]{{\cal{B}}\lpar{#1}\rpar}
\newcommand{\Vverti}[3]{V_{#1}^{#2}\lpar{#3}\rpar}
\newcommand{\Averti}[3]{A_{#1}^{#2}\lpar{#3}\rpar}
\newcommand{\Gverti}[3]{G_{#1}^{#2}\lpar{#3}\rpar}
\newcommand{\Zverti}[3]{Z_{#1}^{#2}\lpar{#3}\rpar}
\newcommand{\Hverti}[2]{H^{#1}\lpar{#2}\rpar}
\newcommand{\Wverti}[3]{W_{#1}^{#2}\lpar{#3}\rpar}
\newcommand{\Cverti}[2]{{\cal{V}}_{#1}^{#2}}
\newcommand{\vverti}[3]{F^{#1}_{#2}\lpar{#3}\rpar}
\newcommand{\averti}[3]{{\overline{F}}^{#1}_{#2}\lpar{#3}\rpar}
\newcommand{\fveone}[1]{f_{#1}}
\newcommand{\fvetri}[3]{f^{#1}_{#2}\lpar{#3}\rpar}
\newcommand{\gvetri}[3]{g^{#1}_{#2}\lpar{#3}\rpar}
\newcommand{\cvetri}[3]{{\cal{F}}^{#1}_{#2}\lpar{#3}\rpar}
\newcommand{\hvetri}[3]{{\hat{\cal{F}}}^{#1}_{#2}\lpar{#3}\rpar}
\newcommand{\avetri}[3]{{\overline{\cal{F}}}^{#1}_{#2}\lpar{#3}\rpar}
\newcommand{\fverti}[2]{F^{#1}_{#2}}
\newcommand{\cverti}[2]{{\cal{F}}_{#1}^{#2}}
\newcommand{\fV}{f_{_{\Vvert}}}
\newcommand{\gA}{g_{_{\Avert}}}
\newcommand{\fVi}[1]{f^{#1}_{_{\Vvert}}}
\newcommand{\seai}[1]{a_{#1}}
\newcommand{\seapi}[1]{a'_{#1}}
\newcommand{\seAi}[2]{A_{#1}^{#2}}
\newcommand{\sewi}[1]{w_{#1}}
\newcommand{\seWi}[1]{W_{#1}}
\newcommand{\seWsi}[1]{W^{*}_{#1}}
\newcommand{\seWti}[2]{W_{#1}^{#2}}
\newcommand{\sewti}[2]{w_{#1}^{#2}}
\newcommand{\seSig}[1]{\Sigma_{#1}\lpar\sla{\pmom}\rpar}
\newcommand{\ww}{{\rm{w}}}
\newcommand{\ew}{{\rm{ew}}}
\newcommand{\ct}{{\rm{ct}}}
\newcommand{\nonSE}{\rm{non-SE}}
\newcommand{\leading}{\rm{\ssL}}
%
% gp: sm_renorm_oneloop
%
\newcommand{\bbff}[1]{{\overline B}_{#1}}
\newcommand{\sW}{p_{_W}}
\newcommand{\sZ}{p_{_Z}}
\newcommand{\ssp}{s_p}
\newcommand{\fW}{f_{_W}}
\newcommand{\fZ}{f_{_Z}}
\newcommand{\subMSB}[1]{{#1}_{\mbox{$\overline{\scriptscriptstyle MS}$}}}
\newcommand{\supMSB}[1]{{#1}^{\mbox{$\overline{\scriptscriptstyle MS}$}}}
\newcommand{\redMSB}{{\mbox{$\overline{\scriptscriptstyle MS}$}}}
\newcommand{\gpbb}{g'_{0}}
\newcommand{\Zconip}[1]{Z'_{#1}}
\newcommand{\bpff}[4]{B'_{#1}\lpar #2;#3,#4\rpar}             % B' form-factor
\newcommand{\xidf}{\xi^2-1}
\newcommand{\tDdr}{1/{\bar{\varepsilon}}}
\newcommand{\cRz}{{\cal R}_{_Z}}
\newcommand{\cRg}{{\cal R}_{\gamma}}
\newcommand{\Sz}{\Sigma_{_Z}}
\newcommand{\alh}{{\hat\alpha}}
\newcommand{\alhz}{\alpha_{\ssZ}}
\newcommand{\Phzg}{{\hat\Pi}_{_{\zb\ab}}}
\newcommand{\fvvert}{F^{\rm vert}_{_V}}
\newcommand{\gavert}{G^{\rm vert}_{_A}}
\newcommand{\bmv}{{\overline m}_{_V}}
\newcommand{\Sgn}{\Sigma_{\gamma\hkn}}
\newcommand{\rmboxd}{{\rm Box}_d\lpar s,t,u;M_1,M_2,M_3,M_4\rpar}
\newcommand{\rmboxc}{{\rm Box}_c\lpar s,t,u;M_1,M_2,M_3,M_4\rpar}
%
% D-functions
%
\newcommand{\Afaci}[1]{A_{#1}}
\newcommand{\Afacis}[1]{A^2_{#1}}
\newcommand{\upar}[1]{u}
\newcommand{\upari}[1]{u_{#1}}
\newcommand{\vpari}[1]{v_{#1}}
\newcommand{\lpari}[1]{l_{#1}}
\newcommand{\Lpari}[1]{l_{#1}}
\newcommand{\Nff}[2]{N^{(#1)}_{#2}}
\newcommand{\Sff}[2]{S^{(#1)}_{#2}}
\newcommand{\sSff}{S}
\newcommand{\etafd}[2]{\eta_d\lpar#1,#2\rpar}
\newcommand{\sigdu}[2]{\sigma_{#1#2}}
\newcommand{\scalc}[4]{c_{0}\lpar #1;#2,#3,#4\rpar}
\newcommand{\scald}[2]{d_{0}\lpar #1,#2\rpar}
\newcommand{\scaldi}[3]{d_{0}^{#1}\lpar #2,#3\rpar}
\newcommand{\pir}[1]{\Pi^{\rm{\ssR}}\lpar #1\rpar}
\newcommand{\sigh}{\sigma_{\rm had}}
\newcommand{\dah}{\Delta\alpha^{(5)}_{\rm had}}
\newcommand{\dat}{\Delta\alpha_{\rm top}}
\newcommand{\Vqed}[3]{V_1^{\rm sub}\lpar#1;#2,#3\rpar}
\newcommand{\thetah}{{\hat\theta}}
\newcommand{\smlon}{\frac{\mlones}{s}}
\newcommand{\lntwo}{\ln 2}
\newcommand{\wmin}{w_{\rm min}}
\newcommand{\kmin}{k_{\rm min}}
\newcommand{\mdls}{\Big|}
\newcommand{\smf}{\frac{\mfs}{s}}
\newcommand{\bint}{\beta_{\rm int}}
\newcommand{\IRv}{V_{_{\rm IR}}}
\newcommand{\IRr}{R_{_{\rm IR}}}
\newcommand{\fssts}{\frac{s^2}{t^2}}
\newcommand{\fssus}{\frac{s^2}{u^2}}
\newcommand{\optM}{1+\frac{t}{M^2}}
\newcommand{\opuM}{1+\frac{u}{M^2}}
\newcommand{\ftM}{\lpar -\frac{t}{M^2}\rpar}
\newcommand{\fuM}{\lpar -\frac{u}{M^2}\rpar}
\newcommand{\omsM}{1-\frac{s}{M^2}}
\newcommand{\xsf}{\sigma_{_{\rm F}}}
\newcommand{\xsb}{\sigma_{_{\rm B}}}
\newcommand{\afb}{A_{_{\rm FB}}}
\newcommand{\rsoft}{\rm soft}
\newcommand{\rms}{\rm s}
\newcommand{\rsmx}{\sqrt{s_{\rm max}}}
\newcommand{\rspm}{\sqrt{s_{\pm}}}
\newcommand{\rsp}{\sqrt{s_{+}}}
\newcommand{\rsm}{\sqrt{s_{-}}}
\newcommand{\sigmx}{\sigma_{\rm max}}
\newcommand{\gG}[2]{G_{#1}^{#2}}
\newcommand{\gacomm}[2]{\lpar #1 - #2\gfd\rpar}
\newcommand{\fcsx}{\frac{1}{\ctwsix}}
\newcommand{\fcq}{\frac{1}{\ctwf}}
\newcommand{\fcs}{\frac{1}{\ctws}}
\newcommand{\affs}[2]{{\cal A}_{#1}\lpar #2\rpar}                   % A
\newcommand{\stwei}{s_{\theta}^8}
\def\mdan{\vspace{1mm}\mpar{\hfil$\downarrow$new\hfil}\vspace{-1mm}
          \ignorespaces}
\def\muan{\vspace{-1mm}\mpar{\hfil$\uparrow$new\hfil}\vspace{1mm}\ignorespaces}
\def\mlan{\vspace{-1mm}\mpar{\hfil$\rightarrow$new\hfil}\vspace{1mm}\ignorespaces}
\def\mnnew{\mpar{\hfil NEWNEW \hfil}\ignorespaces}
%
% db's of libraries
%                  
\newcommand{\boxc}[2]{{\cal{B}}_{#1}^{#2}}
\newcommand{\boxct}[3]{{\cal{B}}_{#1}^{#2}\lpar{#3}\rpar}
\newcommand{\hboxc}[3]{\hat{{\cal{B}}}_{#1}^{#2}\lpar{#3}\rpar}
\newcommand{\vev}{\langle v \rangle}
\newcommand{\vevi}[1]{\langle v_{#1}\rangle}
\newcommand{\vevs}{\langle v^2   \rangle}
\newcommand{\fwfrV}[5]{\Sigma_{_V}\lpar #1,#2,#3;#4,#5 \rpar}
\newcommand{\fwfrS}[7]{\Sigma_{_S}\lpar #1,#2,#3;#4,#5;#6,#7 \rpar}
\newcommand{\fSi}[1]{f^{#1}_{_{\Svert}}}
\newcommand{\fPi}[1]{f^{#1}_{_{\Pvert}}}
\newcommand{\mXs}{m_{_X}}
\newcommand{\mXss}{m^2_{_X}}
\newcommand{\mYs}{M^2_{_Y}}
\newcommand{\xik}{\xi_k}
\newcommand{\xiks}{\xi^2_k}
\newcommand{\mpls}{m^2_+}
\newcommand{\mmis}{m^2_-}
%
%--
\newcommand{\SN}{\Sigma_{_N}}
\newcommand{\SC}{\Sigma_{_C}}
\newcommand{\SPN}{\Sigma'_{_N}}
\newcommand{\PFf}{\Pi^{\fer}_{_F}}
\newcommand{\PFb}{\Pi^{\bos}_{_F}}
\newcommand{\dPZ}{\Delta{\hat\Pi}_{_Z}}
\newcommand{\Sfin}{\Sigma_{_F}}
\newcommand{\Sfir}{\Sigma_{_R}}
\newcommand{\Sfinh}{{\hat\Sigma}_{_F}}
\newcommand{\Sfinf}{\Sigma^{\fer}_{_F}}
\newcommand{\Sfinbh}{\Sigma^{\bos}_{_F}}
\newcommand{\alf}{\alpha^{\fer}}
\newcommand{\alhfz}{\alpha^{\fer}\lpar{\mzs}\rpar}
\newcommand{\alhfs}{\alpha^{\fer}\lpar{\sman}\rpar}
\newcommand{\gfQ}{g^f_{_{Q}}}
\newcommand{\gfL}{g^f_{_{L}}}
\newcommand{\ccf}{\frac{\gbs}{16\,\pi^2}}
\newcommand{\chq}{{\hat c}^4}
\newcommand{\muuq}{m_{u'}}
\newcommand{\muus}{m^2_{u'}}
\newcommand{\mdd}{m_{d'}}
%--
\newcommand{\clf}[2]{\mathrm{Cli}_{_#1}\lpar\displaystyle{#2}\rpar}
\def\stes{\sin^2\theta}
\def\acal{\cal A}
\def\alr{A_{_{\rm{LR}}}}
\newcommand{\barQ}{\overline Q}
\newcommand{\Sptg}{\Sigma'_{_{3Q}}}
\newcommand{\Sptt}{\Sigma'_{_{33}}}
\newcommand{\Ppgg}{\Pi'_{\ph\ph}}
\newcommand{\Pww}{\Pi_{_{\wb\wb}}}
\newcommand{\capV}[2]{{\cal F}^{#2}_{_{#1}}}
%--
\newcommand{\bt}{\beta_t}
\newcommand{\btp}{\beta'_t}
\newcommand{\mhsix}{M^6_{_H}}
\newcommand{\topt}{{\cal T}_{33}}
\newcommand{\topq}{{\cal T}_4}
\newcommand{\Phzgf}{{\hat\Pi}^{\fer}_{_{\zb\ab}}}
\newcommand{\Phzgb}{{\hat\Pi}^{\bos}_{_{\zb\ab}}}
\newcommand{\Sfirh}{{\hat\Sigma}_{_R}}
\newcommand{\Szgh}{{\hat\Sigma}_{_{\zb\ab}}}
\newcommand{\Szghb}{{\hat\Sigma}^{\bos}_{_{\zb\ab}}}
\newcommand{\Szghf}{{\hat\Sigma}^{\fer}_{_{\zb\ab}}}
\newcommand{\Szgb}{\Sigma^{\bos}_{_{\zb\ab}}}
\newcommand{\Szgf}{\Sigma^{\fer}_{_{\zb\ab}}}
%--
\newcommand{\chig}{\chi_{\ph}}
\newcommand{\chiz}{\chi_{\ssZ}}
\newcommand{\Sfih}{{\hat\Sigma}}
\newcommand{\Szzh}{\hat{\Sigma}_{_{\zb\zb}}}
\newcommand{\dPZf}{\Delta{\hat\Pi}^f_{_{\zb}}}
\newcommand{\khZdf}[1]{{\hat\kappa}^{#1}_{_{\zb}}}
\newcommand{\chf}{{\hat c}^4}
%---
%for process sm_ola
\newcommand{\amp}[2]{{\cal{A}}_{_{#1}}^{\rm{#2}}}
%--
\newcommand{\hatvm}[1]{{\hat v}^-_{#1}}
\newcommand{\hatvp}[1]{{\hat v}^+_{#1}}
\newcommand{\hatvpm}[1]{{\hat v}^{\pm}_{#1}}
\newcommand{\kvz}[1]{\kappa^{\zb #1}_{_V}}
%--
\newcommand{\barp}{\overline p}                
\newcommand{\delw}{\Delta_{_{\wb}}}
\newcommand{\bdelw}{{\bar{\Delta}}_{_{\wb}}}
\newcommand{\bdelf}{{\bar{\Delta}}_{\ff}}
\newcommand{\delz}{\Delta_{_\zb}}
\newcommand{\deli}[1]{\Delta\lpar{#1}\rpar}
\newcommand{\hdeli}[1]{{\hat{\Delta}}\lpar{#1}\rpar}
\newcommand{\chizb}{\chi_{_\zb}}
\newcommand{\Swwp}{\Sigma'_{_{\wb\wb}}}
\newcommand{\epph}{\varepsilon'/2}
\newcommand{\sbffp}[1]{B'_{#1}}                    
\newcommand{\epss}{\varepsilon^*}
%--
\newcommand{\Ddrhs}{{\ds\frac{1}{\hat{\varepsilon}^2}}}
\newcommand{\lnmsb}{L_{_\wb}}
\newcommand{\lnsmsb}{L^2_{_\wb}}
\newcommand{\tpni}{\lpar 2\pi\rpar^n\ib}
\newcommand{\tpn}{2^n\,\pi^{n-2}}
%--
\newcommand{\cmf}{M_f}
\newcommand{\cmfs}{M^2_f}
\newcommand{\toDdr}{{\ds\frac{2}{{\bar{\varepsilon}}}}}
\newcommand{\troDdr}{{\ds\frac{3}{{\bar{\varepsilon}}}}}
\newcommand{\totDdr}{{\ds\frac{3}{{2\,\bar{\varepsilon}}}}}
\newcommand{\foDdr}{{\ds\frac{4}{{\bar{\varepsilon}}}}}
\newcommand{\smh}{m_{_H}}
\newcommand{\smhs}{m^2_{_H}}
\newcommand{\Ph}{\Pi_{_\hb}}
\newcommand{\Sphh}{\Sigma'_{_{\hb\hb}}}
\newcommand{\bh}{\beta}
\newcommand{\alsn}{\alpha^{(n_f)}_{_S}}
\newcommand{\smq}{m_q}
\newcommand{\smqp}{m_{q'}}
\newcommand{\shb}{h}
\newcommand{\hab}{A}
\newcommand{\hbpm}{H^{\pm}}
\newcommand{\hbp}{H^{+}}
\newcommand{\hbm}{H^{-}}
\newcommand{\msh}{M_h}
\newcommand{\mha}{M_{_A}}
\newcommand{\mhc}{M_{_{H^{\pm}}}}
\newcommand{\mshs}{M^2_h}
\newcommand{\mhas}{M^2_{_A}}
\newcommand{\barfp}{\overline{f'}}                
\newcommand{\chiii}{{\hat c}^3}
\newcommand{\chiv}{{\hat c}^4}
\newcommand{\chv}{{\hat c}^5}
\newcommand{\chvi}{{\hat c}^6}
\newcommand{\alsvi}{\alpha^{6}_{_S}}
\newcommand{\tww}{t_{_W}}
\newcommand{\ti}{t_{1}}
\newcommand{\tii}{t_{2}}
\newcommand{\tiii}{t_{3}}
\newcommand{\tiv}{t_{4}}
\newcommand{\psla}{\hbox{\rlap/p}}
\newcommand{\qsla}{\hbox{\rlap/q}}
\newcommand{\nsla}{\hbox{\rlap/n}}
\newcommand{\lsla}{\hbox{\rlap/l}}
\newcommand{\msla}{\hbox{\rlap/m}}
\newcommand{\cnsla}{\hbox{\rlap/N}}
\newcommand{\clsla}{\hbox{\rlap/L}}
\newcommand{\cmsla}{\hbox{\rlap/M}}
\newcommand{\blmt}{\lrbr - 3\rrbr}
\newcommand{\blfo}{\lrbr 4 1\rrbr}
\newcommand{\bltp}{\lrbr 2 +\rrbr}
%---------------------------------
% mixed QCD
\newcommand{\clitwo}[1]{{\rm{Li}}_{2}\lpar{#1}\rpar}
\newcommand{\clitri}[1]{{\rm{Li}}_{3}\lpar{#1}\rpar}
% subleadings
\newcommand{\xt}{x_{\ft}}
\newcommand{\zt}{z_{\ft}}
\newcommand{\Ht}{h_{\ft}}
\newcommand{\xts}{x^2_{\ft}}
\newcommand{\zts}{z^2_{\ft}}
\newcommand{\Hts}{h^2_{\ft}}
\newcommand{\ztc}{z^3_{\ft}}
\newcommand{\Htc}{h^3_{\ft}}
\newcommand{\ztq}{z^4_{\ft}}
\newcommand{\Htq}{h^4_{\ft}}
\newcommand{\ztv}{z^5_{\ft}}
\newcommand{\Htv}{h^5_{\ft}}
\newcommand{\ztx}{z^6_{\ft}}
\newcommand{\Htx}{h^6_{\ft}}
\newcommand{\ztz}{z^7_{\ft}}
\newcommand{\Htz}{h^7_{\ft}}
\newcommand{\sht}{\sqrt{\Ht}}
\newcommand{\atan}[1]{{\rm{arctan}}\lpar{#1}\rpar}
\newcommand{\dbff}[3]{{\hat{B}}_{_{{#2}{#3}}}\lpar{#1}\rpar}
\newcommand{\ztbs}{{\bar{z}}^{2}_{\ft}}
\newcommand{\ztb}{{\bar{z}}_{\ft}}
\newcommand{\Htbs}{{\bar{h}}^{2}_{\ft}}
\newcommand{\Htb}{{\bar{h}}_{\ft}}
\newcommand{\Hztb}{{\bar{hz}}_{\ft}}
\newcommand{\Ln}[1]{{\rm{Ln}}\lpar{#1}\rpar}
\newcommand{\Lns}[1]{{\rm{Ln}}^2\lpar{#1}\rpar}
\newcommand{\wt}{w_{\ft}}
\newcommand{\wts}{w^2_{\ft}}
\newcommand{\wtb}{\overline{w}}
%--
\newcommand{\fra}{\frac{1}{2}}
\newcommand{\frb}{\frac{1}{4}}
\newcommand{\frc}{\frac{3}{2}}
\newcommand{\frd}{\frac{3}{4}}
\newcommand{\fre}{\frac{9}{2}}
\newcommand{\frf}{\frac{9}{4}}
\newcommand{\frg}{\frac{5}{4}}
\newcommand{\frh}{\frac{5}{2}}
\newcommand{\fri}{\frac{1}{8}}
\newcommand{\frj}{\frac{7}{4}}
\newcommand{\frl}{\frac{7}{8}}
\newcommand{\Spzzh}{\hat{\Sigma}'_{_{\zb\zb}}}
% Temporary change!!!!!!!!!!
%\newcommand{\sss}{s\sqrt{s}}
\newcommand{\sqs}{\sqrt{s}}
\newcommand{\Rtg}{R_{_{3Q}}}
\newcommand{\Rtt}{R_{_{33}}}
\newcommand{\Rww}{R_{_{\wb\wb}}}
%--
%-- scriptscriptstyle
%--
\newcommand{\ssA}{{\scriptscriptstyle{A}}}
\newcommand{\ssB}{{\scriptscriptstyle{B}}}
\newcommand{\ssC}{{\scriptscriptstyle{C}}}
\newcommand{\ssD}{{\scriptscriptstyle{D}}}
\newcommand{\ssE}{{\scriptscriptstyle{E}}}
\newcommand{\ssF}{{\scriptscriptstyle{F}}}
\newcommand{\ssG}{{\scriptscriptstyle{G}}}
\newcommand{\ssH}{{\scriptscriptstyle{H}}}
\newcommand{\ssI}{{\scriptscriptstyle{I}}}
\newcommand{\ssJ}{{\scriptscriptstyle{J}}}
\newcommand{\ssK}{{\scriptscriptstyle{K}}}
\newcommand{\ssL}{{\scriptscriptstyle{L}}}
\newcommand{\ssM}{{\scriptscriptstyle{M}}}
\newcommand{\ssN}{{\scriptscriptstyle{N}}}
\newcommand{\ssO}{{\scriptscriptstyle{O}}}
\newcommand{\ssP}{{\scriptscriptstyle{P}}}
\newcommand{\ssQ}{{\scriptscriptstyle{Q}}}
\newcommand{\ssR}{{\scriptscriptstyle{R}}}
\newcommand{\ssS}{{\scriptscriptstyle{S}}}
\newcommand{\ssT}{{\scriptscriptstyle{T}}}
\newcommand{\ssU}{{\scriptscriptstyle{U}}}
\newcommand{\ssV}{{\scriptscriptstyle{V}}}
\newcommand{\ssW}{{\scriptscriptstyle{W}}}
\newcommand{\ssX}{{\scriptscriptstyle{X}}}
\newcommand{\ssY}{{\scriptscriptstyle{Y}}}
\newcommand{\ssZ}{{\scriptscriptstyle{Z}}}
\newcommand{\ssWF}{\rm{\scriptscriptstyle{WF}}}
\newcommand{\OLA}{\rm{\scriptscriptstyle{OLA}}}
\newcommand{\QED}{\rm{\scriptscriptstyle{QED}}}
\newcommand{\QCD}{\rm{\scriptscriptstyle{QCD}}}
\newcommand{\EW}{\rm{\scriptscriptstyle{EW}}}
\newcommand{\UV}{\rm{\scriptscriptstyle{UV}}}
\newcommand{\IR}{\rm{\scriptscriptstyle{IR}}}
\newcommand{\OMS}{\rm{\scriptscriptstyle{OMS}}}
\newcommand{\GMS}{\rm{\scriptscriptstyle{GMS}}}
\newcommand{\sszero}{\rm{\scriptscriptstyle{0}}}
%--
\newcommand{\DiagramFermionToBosonFullWithMomenta}[8][70]{
  \vcenter{\hbox{
  \SetScale{0.8}
  \begin{picture}(#1,50)(15,15)
    \put(27,22){$\nearrow$}      
    \put(27,54){$\searrow$}    
    \put(59,29){$\to$}    
    \ArrowLine(25,25)(50,50)      \Text(34,20)[lc]{#6} \Text(11,20)[lc]{#3}
    \ArrowLine(50,50)(25,75)      \Text(34,60)[lc]{#7} \Text(11,60)[lc]{#4}
    \Photon(50,50)(90,50){2}{8}   \Text(80,40)[lc]{#2} \Text(55,33)[ct]{#8}
    \Vertex(50,50){2,5}          \Text(60,48)[cb]{#5} 
    \Vertex(90,50){2}
  \end{picture}}}
  }
\newcommand{\DiagramFermionToBosonPropagator}[4][85]{
  \vcenter{\hbox{
  \SetScale{0.8}
  \begin{picture}(#1,50)(15,15)
    \ArrowLine(25,25)(50,50)
    \ArrowLine(50,50)(25,75)
    \Photon(50,50)(105,50){2}{8}   \Text(90,40)[lc]{#2}
%    \GCirc(50,50){10}{0.5}         \Text(80,48)[cb]{#3}
    \Vertex(50,50){0.5}         \Text(80,48)[cb]{#3}
    \GCirc(82,50){8}{1}            \Text(55,48)[cb]{#4}
    \Vertex(105,50){2}
  \end{picture}}}
  }
\newcommand{\DiagramFermionToBosonEffective}[3][70]{
  \vcenter{\hbox{
  \SetScale{0.8}
  \begin{picture}(#1,50)(15,15)
    \ArrowLine(25,25)(50,50)
    \ArrowLine(50,50)(25,75)
    \Photon(50,50)(90,50){2}{8}   \Text(80,40)[lc]{#2}
    \BBoxc(50,50)(5,5)            \Text(55,48)[cb]{#3}
    \Vertex(90,50){2}
  \end{picture}}}
  }
\newcommand{\DiagramFermionToBosonFull}[3][70]{
  \vcenter{\hbox{
  \SetScale{0.8}
  \begin{picture}(#1,50)(15,15)
    \ArrowLine(25,25)(50,50)
    \ArrowLine(50,50)(25,75)
    \Photon(50,50)(90,50){2}{8}   \Text(80,40)[lc]{#2}
    \Vertex(50,50){2.5}          \Text(60,48)[cb]{#3}
    \Vertex(90,50){2}
  \end{picture}}}
  }
%--
\newcommand{\expgw}{\frac{\gf\mws}{2\srt\,\pi^2}}
\newcommand{\expgz}{\frac{\gf\mzs}{2\srt\,\pi^2}}
\newcommand{\Spww}{\Sigma'_{_{\wb\wb}}}
\newcommand{\shf}{{\hat s}^4}
%--
\newcommand{\acz}{\scff{0}}
\newcommand{\acoo}{\scff{11}}
\newcommand{\acod}{\scff{12}}
\newcommand{\acdo}{\scff{21}}
\newcommand{\acdd}{\scff{22}}
\newcommand{\acdt}{\scff{23}}
\newcommand{\acdq}{\scff{24}}
\newcommand{\acto}{\scff{31}}
\newcommand{\actd}{\scff{32}}
\newcommand{\actt}{\scff{33}}
\newcommand{\actq}{\scff{34}}
\newcommand{\actc}{\scff{35}}
\newcommand{\acts}{\scff{36}}
\newcommand{\acoA}{\scff{1A}}
\newcommand{\acdA}{\scff{2A}}
\newcommand{\acdB}{\scff{2B}}
\newcommand{\acdC}{\scff{2C}}
\newcommand{\acdD}{\scff{2D}}
\newcommand{\actA}{\scff{3A}}
\newcommand{\actB}{\scff{3B}}
\newcommand{\actC}{\scff{3C}}
%--
\newcommand{\ada}{\sdff{0}}
\newcommand{\adb}{\sdff{11}}
\newcommand{\adc}{\sdff{12}}
\newcommand{\add}{\sdff{13}}
\newcommand{\ade}{\sdff{21}}
\newcommand{\adf}{\sdff{22}}
\newcommand{\adg}{\sdff{23}}
\newcommand{\adh}{\sdff{24}}
\newcommand{\adi}{\sdff{25}}
\newcommand{\adj}{\sdff{26}}
\newcommand{\adl}{\sdff{27}}
\newcommand{\adm}{\sdff{31}}
\newcommand{\adn}{\sdff{32}}
\newcommand{\ado}{\sdff{33}}
\newcommand{\adp}{\sdff{34}}
\newcommand{\adq}{\sdff{35}}
\newcommand{\adr}{\sdff{36}}
\newcommand{\ads}{\sdff{37}}
\newcommand{\adt}{\sdff{38}}
\newcommand{\adu}{\sdff{39}}
\newcommand{\adw}{\sdff{310}}
\newcommand{\adv}{\sdff{311}}
\newcommand{\ady}{\sdff{312}}
\newcommand{\adz}{\sdff{313}}
%--
\newcommand{\admt}{\frac{\tman}{\sman}}
\newcommand{\admu}{\frac{\uman}{\sman}}
\newcommand{\frm}{\frac{3}{8}}
\newcommand{\frn}{\frac{5}{8}}
\newcommand{\fro}{\frac{15}{8}}
\newcommand{\frp}{\frac{3}{16}}
\newcommand{\frq}{\frac{5}{16}}
\newcommand{\frr}{\frac{1}{16}}
\newcommand{\frs}{\frac{7}{2}}
\newcommand{\frt}{\frac{7}{16}}
\newcommand{\fru}{\frac{1}{3}}
\newcommand{\frw}{\frac{2}{3}}
\newcommand{\frz}{\frac{4}{3}}
\newcommand{\fry}{\frac{13}{3}}
\newcommand{\fraa}{\frac{11}{4}}
%--
\newcommand{\bee}{\beta_{e}}
\newcommand{\beW}{\beta_{_\wb}}
\newcommand{\beWs}{\beta^2_{_\wb}}
\newcommand{\etaW}{\eta_{_\wb}}
%--
\newcommand{\toDdrh}{{\ds\frac{2}{{\hat{\varepsilon}}}}}
\newcommand{\bqas}{\begin{eqnarray*}}
\newcommand{\eqas}{\end{eqnarray*}}
%--
\newcommand{\mhcub}{M^3_{_H}}
%--
\newcommand{\adComA}{\sdff{A}}
\newcommand{\adComB}{\sdff{B}}
\newcommand{\adComC}{\sdff{C}}
\newcommand{\adComD}{\sdff{D}}
\newcommand{\adComE}{\sdff{E}}
\newcommand{\adComF}{\sdff{F}}
\newcommand{\adComG}{\sdff{G}}
\newcommand{\adComH}{\sdff{H}}
\newcommand{\adComI}{\sdff{I}}
\newcommand{\adComJ}{\sdff{J}}
\newcommand{\adComL}{\sdff{L}}
\newcommand{\adComM}{\sdff{M}}
\newcommand{\adComN}{\sdff{N}}
\newcommand{\adComO}{\sdff{O}}
\newcommand{\adComP}{\sdff{P}}
\newcommand{\adComQ}{\sdff{Q}}
\newcommand{\adComR}{\sdff{R}}
\newcommand{\adComS}{\sdff{S}}
\newcommand{\adComT}{\sdff{T}}
\newcommand{\adComU}{\sdff{U}}
%--
\newcommand{\adComAc}{\sdff{A}^c}
\newcommand{\adComBc}{\sdff{B}^c}
\newcommand{\adComCc}{\sdff{C}^c}
\newcommand{\adComDc}{\sdff{D}^c}
\newcommand{\adComEc}{\sdff{E}^c}
\newcommand{\adComFc}{\sdff{F}^c}
\newcommand{\adComGc}{\sdff{G}^c}
\newcommand{\adComHc}{\sdff{H}^c}
\newcommand{\adComIc}{\sdff{I}^c}
\newcommand{\adComJc}{\sdff{J}^c}
\newcommand{\adComLc}{\sdff{L}^c}
\newcommand{\adComMc}{\sdff{M}^c}
\newcommand{\adComNc}{\sdff{N}^c}
\newcommand{\adComOc}{\sdff{O}^c}
\newcommand{\adComPc}{\sdff{P}^c}
\newcommand{\adComQc}{\sdff{Q}^c}
\newcommand{\adComRc}{\sdff{R}^c}
\newcommand{\adComSc}{\sdff{S}^c}
\newcommand{\adComTc}{\sdff{T}^c}
\newcommand{\adComUc}{\sdff{U}^c}
%--
\newcommand{\adComAf}{\sdff{A}^f}
\newcommand{\adComBf}{\sdff{B}^f}
\newcommand{\adComCf}{\sdff{F}^f}
\newcommand{\adComDf}{\sdff{D}^f}
\newcommand{\adComEf}{\sdff{E}^f}
\newcommand{\adComFf}{\sdff{F}^f}
\newcommand{\adComGf}{\sdff{G}^f}
\newcommand{\adComHf}{\sdff{H}^f}
\newcommand{\adComIf}{\sdff{I}^f}
\newcommand{\adComJf}{\sdff{J}^f}
\newcommand{\adComLf}{\sdff{L}^f}
\newcommand{\adComMf}{\sdff{M}^f}
\newcommand{\adComNf}{\sdff{N}^f}
\newcommand{\adComOf}{\sdff{O}^f}
\newcommand{\adComPf}{\sdff{P}^f}
\newcommand{\adComQf}{\sdff{Q}^f}
\newcommand{\adComRf}{\sdff{R}^f}
\newcommand{\adComSf}{\sdff{S}^f}
\newcommand{\adComTf}{\sdff{T}^f}
\newcommand{\adComUf}{\sdff{U}^f}
%--
\newcommand{\adComBfc}{\sdff{B}^{fc}} 
\newcommand{\adComCfco}{\sdff{C}^{fc1}}
\newcommand{\adComCfcd}{\sdff{C}^{fc2}} 
\newcommand{\adComCfct}{\sdff{C}^{fc3}} 
\newcommand{\adComDfc}{\sdff{D}^{fc}}
\newcommand{\adComEfc}{\sdff{E}^{fc}}
\newcommand{\adComFfc}{\sdff{F}^{fc}}
\newcommand{\adComGfc}{\sdff{G}^{fc}}
\newcommand{\adComHfc}{\sdff{H}^{fc}}
\newcommand{\adComLfc}{\sdff{L}^{fc}}
%--
\newcommand{\afba}[1]{A^{#1}_{_{\rm FB}}}
\newcommand{\alra}[1]{A^{#1}_{_{\rm LR}}}
%--
\newcommand{\adComAt}{\sdff{A}^t}
\newcommand{\adComBt}{\sdff{B}^t}
\newcommand{\adComCt}{\sdff{T}^t}
\newcommand{\adComDt}{\sdff{D}^t}
\newcommand{\adComEt}{\sdff{E}^t}
\newcommand{\adComFt}{\sdff{T}^t}
\newcommand{\adComGt}{\sdff{G}^t}
\newcommand{\adComHt}{\sdff{H}^t}
\newcommand{\adComIt}{\sdff{I}^t}
\newcommand{\adComJt}{\sdff{J}^t}
\newcommand{\adComLt}{\sdff{L}^t}
\newcommand{\adComMt}{\sdff{M}^t}
\newcommand{\adComNt}{\sdff{N}^t}
\newcommand{\adComOt}{\sdff{O}^t}
\newcommand{\adComPt}{\sdff{P}^t}
\newcommand{\adComQt}{\sdff{Q}^t}
\newcommand{\adComRt}{\sdff{R}^t}
\newcommand{\adComSt}{\sdff{S}^t}
\newcommand{\adComTt}{\sdff{T}^t}
\newcommand{\adComUt}{\sdff{U}^t}
%--
\newcommand{\adComAtt}{\sdff{A}^{\tau}}
\newcommand{\adComBtt}{\sdff{B}^{\tau}}
\newcommand{\adComCtt}{\sdff{T}^{\tau}}
\newcommand{\adComDtt}{\sdff{D}^{\tau}}
\newcommand{\adComEtt}{\sdff{E}^{\tau}}
\newcommand{\adComFtt}{\sdff{T}^{\tau}}
\newcommand{\adComGtt}{\sdff{G}^{\tau}}
\newcommand{\adComHtt}{\sdff{H}^{\tau}}
\newcommand{\adComItt}{\sdff{I}^{\tau}}
\newcommand{\adComJtt}{\sdff{J}^{\tau}}
\newcommand{\adComLtt}{\sdff{L}^{\tau}}
\newcommand{\adComMtt}{\sdff{M}^{\tau}}
\newcommand{\adComNtt}{\sdff{N}^{\tau}}
\newcommand{\adComOtt}{\sdff{O}^{\tau}}
\newcommand{\adComPtt}{\sdff{P}^{\tau}}
\newcommand{\adComQtt}{\sdff{Q}^{\tau}}
\newcommand{\adComRtt}{\sdff{R}^{\tau}}
\newcommand{\adComStt}{\sdff{S}^{\tau}}
\newcommand{\adComTtt}{\sdff{T}^{\tau}}
\newcommand{\adComUtt}{\sdff{U}^{\tau}}
%--
\newcommand{\etavz}[1]{\eta^{\zb #1}_{_V}}
%--
\newcommand{\phanst}{$\hphantom{\sigma^{s+t}\ }$}
\newcommand{\phanat}{$\hphantom{A_{FB}^{s+t}\ }$}
\newcommand{\phanss}{$\hphantom{\sigma^{s}\ }$}
\newcommand{\phanas}{$\hphantom{A_{FB}^{s}\ }$} 
\newcommand{\pbb}{\,\mbox{\bf pb}}
\newcommand{\pe}{\,\%\:}
\newcommand{\pc}{\,\%}
%--
\newcommand{\temiv}{10^{-4}}
\newcommand{\temv}{10^{-5}}
\newcommand{\temvi}{10^{-6}}
\newcommand{\di}[1]{d_{#1}}
%--
\newcommand{\delip}[1]{\Delta_+\lpar{#1}\rpar}
\newcommand{\propbb}[5]{{{#1}\over {\lpar #2^2 + #3 - \ib\varepsilon\rpar
\lpar\lpar #4\rpar^2 + #5 -\ib\varepsilon\rpar}}}
\newcommand{\cfft}[5]{C_{#1}\lpar #2;#3,#4,#5\rpar}    
\newcommand{\ppl}[1]{p_{+{#1}}}
\newcommand{\pmi}[1]{p_{-{#1}}}
%--
\newcommand{\bpox}{\beta^2_{\xi}}
\newcommand{\bffdiff}[5]{B_{\rm d}\lpar #1;#2,#3;#4,#5\rpar}             
\newcommand{\cffdiff}[7]{C_{\rm d}\lpar #1;#2,#3,#4;#5,#6,#7\rpar}    
\newcommand{\affdiff}[2]{A_{\rm d}\lpar #1;#2\rpar}             
\newcommand{\Dqf}{\Delta\qf}
\newcommand{\bposx}{\beta^4_{\xi}}
\newcommand{\svverti}[3]{f^{#1}_{#2}\lpar{#3}\rpar}
\newcommand{\Mods}{\mbox{$M^2_{12}$}}
\newcommand{\Mots}{\mbox{$M^2_{13}$}}
\newcommand{\Motq}{\mbox{$M^4_{13}$}}
\newcommand{\Mdts}{\mbox{$M^2_{23}$}}
\newcommand{\Mdos}{\mbox{$M^2_{21}$}}
\newcommand{\Mtds}{\mbox{$M^2_{32}$}}
\newcommand{\dffpt}[3]{D_{#1}\lpar #2,#3;}           
\newcommand{\quu}{Q_{uu}}
\newcommand{\qdd}{Q_{dd}}
\newcommand{\qud}{Q_{ud}}
\newcommand{\qdu}{Q_{du}}
%--
\newcommand{\msPj}[6]{\Lambda^{#1#2#3}_{#4#5#6}}
%--
\newcommand{\bdiff}[4]{B_{\rm d}\lpar #1,#2;#3,#4\rpar}             
\newcommand{\bdifff}[7]{B_{\rm d}\lpar #1;#2;#3;#4,#5;#6,#7\rpar}             
\newcommand{\adiff}[3]{A_{\rm d}\lpar #1;#2;#3\rpar}  
%--
\newcommand{\aw}{a_{_\wb}}
\newcommand{\az}{a_{_\zb}}
%--
\newcommand{\sct}[1]{sect.~\ref{#1}}
%--
\newcommand{\dreim}[1]{\varepsilon^{\rm M}_{#1}}
\newcommand{\drem}{\varepsilon^{\rm M}}
%--
\newcommand{\hcapV}[2]{{\hat{\cal F}}^{#2}_{_{#1}}}
%--
\newcommand{\swww}{{\scriptscriptstyle \wb\wb\wb}}
\newcommand{\szhz}{{\scriptscriptstyle \zb\hb\zb}}
\newcommand{\shzh}{{\scriptscriptstyle \hb\zb\hb}}
\newcommand{\bwith}[3]{\beta^{#3}_{#1}\lpar #2\rpar}
\newcommand{\Shhh}{{\hat\Sigma}_{_{\hb\hb}}}
\newcommand{\Sphhh}{{\hat\Sigma}'_{_{\hb\hb}}}
%--
\newcommand{\seWilc}[1]{w_{#1}}
\newcommand{\seWtilc}[2]{w_{#1}^{#2}}
%--
\newcommand{\eilc}{\gamma}
\newcommand{\eilcs}{\gamma^2}
\newcommand{\eilcc}{\gamma^3}
\newcommand{\eilcb}{{\overline{\gamma}}}
\newcommand{\eilcbs}{{\overline{\gamma}^2}}
%--
\newcommand{\Sttww}{\Sigma_{_{33;\wb\wb}}}
\newcommand{\bSttww}{{\overline\Sigma}_{_{33;\wb\wb}}}
%--
\newcommand{\Pggtg}{\Pi_{\ph\ph;3Q}}
\def\rmL{{\rm L}}
\def\rmT{{\rm T}}
\def\rmM{{\rm M}}
\newcommand{\hsm}{\hspace{-0.4mm}}
\newcommand{\hsmm}{\hspace{-0.2mm}}
\def\negs{\hspace{-0.26in}}
\def\negss{\hspace{-0.18in}}
\def\negsss{\hspace{-0.09in}}
\def\scee{{\mbox{\lowercase{$\fep\fem$}}}}
\def\scffb{{\mbox{\lowercase{$\ff\barf$}}}}
\def\scqq{{\mbox{\lowercase{$\fq\barq$}}}}
\def\scln{{\mbox{\lowercase{$\fl\fnu$}}}}
%----------------
%--
%-- Bibliography
%--
%----------------------------------------------------------
\def\app#1#2 {{\it Acta. Phys. Pol.} {\bf#1},#2}
\def\cpc#1#2 {{\it Computer Phys. Comm.} {\bf#1},#2}
\def\np#1#2 {{\it Nucl. Phys.} {\bf#1},#2}
\def\pl#1#2 {{\it Phys. Lett.} {\bf#1},#2}
\def\prep#1#2 {{\it Phys. Rep.} {\bf#1},#2}
\def\prev#1#2 {{\it Phys. Rev.} {\bf#1},#2}
\def\prl#1#2 {{\it Phys. Rev. Lett.} {\bf#1},#2}
\def\zp#1#2 {{\it Zeit. Phys.} {\bf#1},#2}
\def\sptp#1#2 {{\it Suppl. Prog. Theor. Phys.} {\bf#1},#2}
\def\mpl#1#2 {{\it Modern Phys. Lett.} {\bf#1},#2}
\def\jetp#1#2 {{\it Sov. Phys. JETP} {\bf#1},#2}
\def\fpj#1#2 {{\it Fortschr. Phys.} {\bf#1},#2}
\def\afp#1#2 {{\it Acta.Phys. Polon.} {\bf#1},#2}
\def\err#1#2 {{\it Erratum} {\bf#1},#2}
\def\ijmp#1#2 {{\it Int. J. Mod. Phys} {\bf#1},#2}
\def\nc#1#2 {{\it Nuovo Cimento} {\bf#1},#2}
\def\ap#1#2 {{\it Ann. Phys.} {\bf#1},#2}
\def\cmp#1#2 {{\it Comm. Math. Phys.} {\bf#1},#2}
\def\el#1#2 {{\it Europhys. Lett.} {\bf#1},#2}
\def\hpa#1#2 {{\it Helv. Phys. Acta} {\bf#1},#2}
\def\yf#1#2 {{\it Yad. Fiz.} {\bf#1},#2}
\def\nim#1#2 {{\it Nucl. Instrum. Meth.} {\bf#1},#2}
\def\spz#1#2 {{\it Sov. Pisma Zhetf} {\bf#1},#2}
\def\jetpl#1#2 {{\it JETP Lett.} {\bf#1},#2}
\def\sjnp#1#2 {{\it Sov. J. Nucl. Phys.} {\bf#1},#2}
\def\ptp#1#2 {{\it Progr. Theor. Phys. (Kyoto)} {\bf#1},#2}
\def\rmp#1#2  {{\it Rev. Mod. Phys.} {\bf#1},#2}
\def\zhetf#1#2 {{\it ZhETF} {\bf#1},#2}
\def\prs#1#2 {{\it Proc. Roy. Soc.} {\bf#1},#2}
\def\phys#1#2 {{\it Physica} {\bf#1},#2}
\def\itetal{{\it et al.}}
%---------------------------
% additions from zf_defs.tex
%---------------------------
\newcommand{\dalpha}{\Delta\alpha}
\newcommand{\drho}{\Delta\rho}
\newcommand{\drhov}{\delta\rho}
\newcommand{\dkapv}{\delta\kappa}
\newcommand{\drhovh}{\delta{\hat{\rho}}}
\newcommand{\drhovb}{\delta{\hat{\rho}}}
\newcommand{\sss}[1]{\scriptscriptstyle{#1}}
\newcommand{\epsb}{\bar\varepsilon}
\newcommand{\gll}{\Gamma_{\fl}}
\newcommand{\gqq}{\Gamma_{\fq}}
\newcommand{\Imsi}[1]{I^2_{#1}}
\newcommand{\reni}[1]{R_{#1}}
\newcommand{\renis}[1]{R^2_{#1}}
\newcommand{\sreni}[1]{\sqrt{R_{#1}}}
\newcommand{\dalphav}{\Delta\alpha^{(5)}(\mzs)}
\newcommand{\Rvaz}[1]{g^{#1}_{\sss{\zb}}}
\newcommand{\rvab}[1]{{\bar{g}}_{#1}}
\newcommand{\rvabs}[1]{{\bar{g}}^2_{#1}}
\newcommand{\rab }[1]{{\bar{a}}_{#1}}
\newcommand{\rvb }[1]{{\bar{v}}_{#1}}
\newcommand{\rabs}[1]{{\bar{a}}^2_{#1}}
\newcommand{\rva }[1]{g_{#1}}
\newcommand{\rvas}[1]{g^2_{#1}}
\newcommand{\Rva }[1]{G_{#1}}
\newcommand{\Rvac}[1]{G^{*}_{#1}}
\newcommand{\Rvah }[1]{{\hat{G}}_{#1}}
\newcommand{\Rvahc}[1]{{\hat{G}}^{*}_{#1}}
\newcommand{\Rvas}[1]{G^2_{#1}}
\newcommand{\rhobi} [1]{{\bar{\rho}}_{#1}}
\newcommand{\kappai}[1]{\kappa_{#1}}
\newcommand{\rhois}[1]{\rho^2_{#1}}
\newcommand{\rhohi} [1]{{\hat{\rho}}_{#1}}
\newcommand{\rhobpi}[1]{{\bar{\rho}}'_{#1}}
\newcommand{\dd}{{\mathrm d}}
\newcommand{\matrm}[1]{\mbox{\scriptsize #1}}
\newcommand{\vecc}[1]{\mbox{\boldmath $#1$}}
\newcommand{\bm}[1]{\mbox{\boldmath $#1$}}
%-----------------------------------------
% Lessons
\newcommand{\Kx}{{x}}
\newcommand{\KxT}{{x^{\ssT}}}
\newcommand{\Kxi}{{x}_{_K}}
\newcommand{\KxiT}{{x}^{\ssT}_{_K}}
\newcommand{\Kvec}[1]{{#1}}
\newcommand{\KvecT}[1]{{#1}^{\ssT}}
\newcommand{\Pmat}{\tilde{P}}
\newcommand{\Pmati}{\tilde{P}^{-1}}
\newcommand{\detg}[1]{\mbox{det}_{#1}}
\newcommand{\mtl}{ m_t }
\newcommand{\mzl}{ M_{\sss Z}}
\newcommand{\mzf}{ M_{\sss Z}^4 }
\newcommand{\lpstnt}{ {\hat L}_0(\mtl,0,\mtl)    }
\newcommand{\lpstzn}{ {\hat L}_0(\mtl,\mzl,0)    }
\newcommand{\lpstzs}{ {\hat L}_0(\mtl,\mzl,0)    }
\newcommand{\lpttzt}{ {\hat L}_0(\mtl,\mzl,\mtl) }
\newcommand{\plstnt}{ {\hat L}_1(\mtl,0,\mtl)    }
\newcommand{\plstzn}{ {\hat L}_1(\mtl,\mzl,0)    }
\newcommand{\plstzt}{ {\hat L}_1(\mtl,\mzl,\mtl) }
\newcommand{\plttzt}{ {\hat L}_1(\mtl,\mzl,\mtl) }

\newcommand{\Lnmuzm}{L_{\tHs}(\mzl)}
\newcommand{\Lnmutm}{L_{\tHs}(\mtl)}

\newcommand{\vts}{v^2_t}
\newcommand{\vt }{v_t}
\newcommand{\at }{a_t}
\newcommand{\ats}{a^2_t}

\newcommand{\DGrt  }{\Delta}
 
\newcommand{\nll }{\nonumber \\}
\newcommand{\Simplexl }{\int_0^1 dx }
\newcommand{\Simplexll}{\int_0^1 dx \int_0^{1-x} dy}
\newcommand{\Poly}[1]{P_{x,y}^{#1}}
\newcommand{\DettI}{\ds \frac{1}{\Delta}}

\newcommand{\xvarl}{{\bf x}}
\newcommand{\yvarl}{{\bf y}}

\newcommand{\Pxy}{P_{x,y}} 
\newcommand{\NDetti}{k_3(\mtl,\mzl,\mtl)}
\newcommand{\tHmus}{\mu^2}
\newcommand{\NDstnt}{k_2(\mtl,0,\mtl)}
\newcommand{\NDstzn}{k_2(\mtl,\mzl,0)}
\newcommand{\NDstzns}{k^2_2(\mtl,\mzl,0)}
\newcommand{\NDstzt}{k_2(\mzl,0,\mtl)}
\newcommand{\NDttzt}{k_3(\mtl,\mzl,\mtl)}
\newcommand{\NDttzts}{k^2_3(\mtl,\mzl,\mtl)}
\newcommand{\rtzs}{r^2_{tz}}
\newcommand{\st}{s_{t}}
\newcommand{\mzt}{M^2_{{\sss{Z}}t}}
\newcommand{\vvertil}[3]{F^{#1}_{#2}\lpar{#3}\rpar}
\newcommand{\cvetril}[3]{{\cal{F}}^{#1}_{#2}\lpar{#3}\rpar}
\newcommand{\cvertil}[3]{{\cal{F}}^{#1}_{#2}\lpar{#3}\rpar}
\newcommand{\gspi}{\frac{g^2}{16\pi^2}}
\newcommand{\tcit}{I^{(3)}_t}
\newcommand{\rtw }{r_{t{\sss W}}}
\newcommand{\pole}{\dlt}
\newcommand{\dlt}{\displaystyle{\frac{1}{\epsb}}}
\newcommand{\au}{a_t}
\newcommand{\vu}{v_t}
\newcommand{\ruz}{r_{t{\sss Z}}}
\newcommand{\rtz}{r_{t{\sss Z}}}
\newcommand{\ruw}{r_{t{\sss W}}}
\newcommand{\ruh}{r_{t{\sss H}}}
\newcommand{\qum}{\left|Q_{t}\right|}
\newcommand{\Lmmt}{L_\mu(\mts)}
\newcommand{\Lmmz}{L_\mu(\mzs)}
\newcommand{\Lmmh}{L_\mu(\mhs)}
\newcommand{\Lmmw}{L_\mu(\mws)}
\newcommand{\detfi}[1]{\Delta_4\lpar #1      \rpar}
\newcommand{\detti}[3]{\Delta_3\lpar #1,#2,#3\rpar}
\newcommand{\Longab}[3]{L_{ab} \lpar #1,#2,#3\rpar}        
%-------------
\thispagestyle{empty}

{\flushright 
\begin{tabular}{cr}
\hspace{9.5cm} &January 26, 2001\\
               &hep-ph/0101295
\end{tabular}
}

\vspace{1cm}

\begin{center}
{\LARGE\bf 12 Years of Precision Calculations for 
\vspace{5mm}

LEP. What's Next?}

\vspace{2.5cm}

{\Large Dmitri Bardin}
\vspace{5mm}

{\large\it Laboratory of Nuclear Problems, JINR, Dubna, Russia}

\vspace{2cm}

{\large\bf Abstract} 

\vspace{2mm}

\end{center}

\noindent I shortly review time period of twelve years, 1989-2000, which was 
devoted to a theoretical support of experiments at LEP and SLC at $Z$ resonance
and discuss several directions of possible future work in the field of precision
theoretical calculations for experiments at future colliders.

\vspace{3cm}

\begin{center}

Presented at the

{\large

50 Years of Electroweak Physics

A symposium in honor of Professor

Alberto Sirlin's 70th Birthday
\vspace*{1mm}

October 27-28, 2000
}
\end{center}
\newpage
\setcounter{page}{1}
\centerline{\large\bf 12 Years of Precision Calculations for LEP. What's Next?}
\vspace*{2mm}

\centerline{\bf Dmitri Bardin}
\vspace*{2mm}

\centerline{\it  Laboratory of Nuclear Problems, JINR, Dubna, Russia}
%---------------------
\section{Introduction}
%---------------------
Exiting 50 years of foundation of Precision EW Physics were perfectly reviewed
in several brilliant talks at this Symposium~\cite{fourtalks}. I hope, our
community would share my opinion that after 12 years of excellent work of LEP 
we have full rights to say that a new scientific discipline has been born,
Precision High-Energy Physics, {\bf PHEP}, consisting of:
\vspace*{-2mm}
\begin{itemize}
\item[$-$] experimental measurements themselves at per mill precision level;
\vspace*{-3mm} 
\item[$-$] supporting theoretical calculations with even better precision. 
\vspace*{-2mm}
\end{itemize}
Me and my colleagues were involved into this glorious period with {\tt ZFITTER} 
project. 
Actually, {\tt ZFITTER} project has been started 25 years ago, in 1976, when 
we wrote our papers on {\em on-mass-shell renormalization scheme}. Formally,
first {\tt ZFITTER} was born within CERN workshop ``Z Physics at LEP 1'' in 
1989. 
As an important intermediate step I would like to mention CERN workshop 
1994--1995, ``Precision calculations for Z resonance'', 
which I had pleasure 
to run and to have Alberto among our contributors. In 1999 we have published 
a long write-up of {\tt ZFITTER v6.21}, which recently appeared in 
CPC~\cite{Bardin:1999yd}.
When I say ``12 Years of Precision Calculations for LEP'', I mean that during 
all lifetime of LEP, {\bf 1989--2000}, {\tt ZFITTER} was upgraded and supported
in ADLO, SLC and LEPEWWG. We would like to see {\bf the end} of {\tt ZFITTER} 
project in the Y2K and, therefore, a very natural question arises: 
{\bf What's next?}

Meantime, me and Giampiero Passarino in {\bf 1997-1998} wrote a monstrous 
book: ``The Standard Model in the Making'', which appeared in 1999 in 
Oxford University Press~\cite{Bardin:1999ak}. 
It has some value for this talk.

{\tt ZFITTER} uses an one-loop core and general environment based on our own 
formulae, and incorporates all the world results for higher order QED, QCD and 
EW 
Radiative Corrections (EWRC). As an input it uses the so-called {\bf LEP1, IPS}
-- Input Parameter Set, i.e. 5 parameters: 
\vspace*{-1.5mm}
%--
\bq
\dalpha^{(5)}_{\had}(\mzs),\qquad\als\lpar\mzs\rpar,\qquad{\mt}\,,
\qquad\mz\,,\qquad{\bf\mh}\,.
\eq 
%--
\vspace*{-5.5mm}

\noindent
$\mz$ is measured very precise at LEP1 and for the first three
para\-meters a rich information is available from the other measurements.
Therefore, we are approaching one-parameter fit with ${\bf \mh}$ being the only
parameter. Results of such a fit are being presented in famous 
{\bf{\em Blue Band}} 
figure, derived with the aid of {\tt TOPAZ0 \& ZFITTER} codes.

{\tt ZFITTER} incorporates all known in the literature {QED} RC up to
$\ord{(\alpha L)^3}$ (where $L=\ln\sman/\mes-1=23\;\mbox{at}\;\sman=\mzs$ and,
therefore, effective QED coupling is quite large: $\alpha L = 0.169$); 
{QCD} RC up to $\ord{\als^3}$; and EWRC up to 
$\ord{\alpha^2}$, however, for the latter only {\em leading} and 
{\em subleading} RCs are known ({\it Degrassi, Gambino,
Sirlin et al., 1996--1997}).

In several cases perturbative calculations are {\it saturated} like it takes
place for leptonic contribution to the running QED coupling, 
$\dalpha_{\lep}(\sman)$, which is presently known
up to three loops (2-loops {\it K{\"a}llen, 1955}; 
3-loops {\it Steinhauser, 1998}):
\vspace*{-1mm}
%--
\bq
\dalpha_{\lep}= 314.97637\cdot 10^{-4} 
 = \bigl[ 314.18942_{\rm 1-loop}
          + 0.77616_{\rm 2-loop}
          + 0.01079_{\rm 3-loop} \bigr] \cdot 10^{-4}.
\eq
%--
\vspace*{-6mm}

\noindent
The last known term is small enough to be used as an {\em estimator}
of theoretical uncertainty. Typical scales of the problem:
$\mbox{LEP1,2 energies}\;\sqs = \mz-200\;\mbox{GeV}\;,\mw\,,\mz\,,\mt\,,\mh\,,$
all are of the same order and calculations must be {\bf complete} and  
one-loop calculations were complete!
For two-loops, the notion of $\mts$ {\em enhanced terms} was introduced.
Only terms $\ord{\gf\mtq}$ and $\ord{\gf\mts\mzs}$ are known.
Since likely, ~$100\;\mbox{GeV}\leq\mh\leq 250\;\mbox{GeV},$
popular expansions in $\mhs/\mts$ and $\mts/\mhs$ have bad convergence.
One may say that 
{\bf complete two-loop EWRC were welcome for LEP1 but our community did not
deliver them!}

In this talk I will tell about three directions of possible future work 
in which we were digging in the Y2K. The first one we call:
%--------------------------
\section{Book ``heritage''}
%--------------------------
While working on the book, me and Giampiero Passarino wrote 
dozens of book supporting {\tt form} codes. Later on
{\em an idea was erupted to collect, order, unify and 
upgrade these codes up to the level of a ``computer system''}:
to which we gave the name {\tt CalcPHEP} --- 
{\em 'Calc'ulus of Precision High Energy Physics}.

This system is being realized at a web site: {\tt brg.jinr.ru}.
Presently we do not maintain any author list, we say that we have people who
contributed to it:
{\em G.Passarino, DB, L.Kalinovskaya, P.Christova, G.Nanava and A.Andonov}
\cite{CalcPHEP:2000}.

After completion of R\&D phase of the project, presently the following options 
are available at the site: 
\vspace*{-2mm}
\begin{enumerate}
\item generation and reduction to the scalar PV functions of one-loop 
Feynman diagrams (FD) for {\bf all} SM $1\to 2$ decays and $2f\to 2f$ 
processes (in $R_{\xi}$ gauge, QCD included);
\vspace*{-3mm}
\item computation of one-loop scalar form factors for decays 
$Z(H)\to f\bar{f}$ and $W\to f_1\bar{f}_2$;
\end{enumerate}
\vspace*{-2mm}
The system was already used for calculation of EWRC to 
$\fep\fem\to\ft\bart\,$ (as described in Section 3).

In case if the project will be approved by scientific community
(I mean, it will be decided that it is worth doing, since there is certain 
competition with FeynArts, see, for instance~\cite{Hahn:2000jm}), 
during next two years (2001--2002) it would be feasible to realize next steps: 
\vspace*{-3mm}

\begin{enumerate}
\item extension of computation of one-loop scalar form factors for all
SM $1\to 2$ decays and $2f\to 2f$ processes of expe\-rimental interest 
({\bf REI criterion} -- {\it Reactions of Experimental Interest}. 
By this we mean
that in the initial state could be only $\fepm,\ph$, muons, neutrinos and 
partons).
\vspace*{-7mm}
\item extension of availability of generation and reduction of FD 
for all $2\to 2$ processes;
\vspace*{-3mm}
\item realization of the step: 
\underline{from form factors to helicity amplitudes};
\vspace*{-3mm}
\item realization of the step:
\underline{from amplitudes to realistic observables};
\vspace*{-3mm}
\item solution of the problem of automatic generation of codes for 
numerical calculation of form factors, amplitudes and observables;
\vspace*{-3mm}
\item creation of codes for calculation of one-loop amplitudes 
for SM REI $2\to 3$ processes;
\vspace*{-3mm}
\item user support of {\tt CalcPHEP} system;
creation of user-friendly environment on the site.
\end{enumerate}
%--------------

%--------------------------------------------------------------------------
\section{New calculation for $\fep\fem\hspace*{-1.5mm}\to\ft\bar{t}$}
%--------------------------------------------------------------------------
The process $e^+e^-\to\ft\bar{t}$ is studied already about ten years 
in connection with experiments at future linear colliders, see
for instance recent review~\cite{Beneke:2000hk}.

Actually, it is a six-fermion process, however
the cross-section 
$\sigma(\fep\fem\to\ft\bar{t})$ with tops on-mass-shell is an ingredient 
in various app\-roaches like DPA~\cite{Denner:2000bj}, 
or the so-called Modified Perturbation Theory (MPT)~\cite{Nekrasov:1999cj}.

Recently we completed a new calculation of the electroweak part of the 
amplitude of $e^+e^-\to\ft\bar{t}$ process~\cite{Bardin:2000kn}
in two gauges, $\Rxi$ and unitary ones.
%This calculation is a first step of the project {\tt topfit}~\cite{topfit:2000},
%which we started in Y2K after finishing {\tt ZFITTER}~\cite{Bardin:1999yd}. 
%So, in a way {\tt topfit} is a successor of {\tt ZFITTER},
%however, our goal is to create a new code from scratch.

There were many studies of $e^+e^-\to\ft\bar{t}$ process in 
$\xi=1$ gauge, see for example \cite{Beenakker:1991ca} and \cite{Driesen:1996tn}.
%however, we are not aware of existence of calculations in $\Rxi$ and unitary gauges.

The {\bf purposes} of this new study are: 1) to explicitly control gauge 
invariance in $\Rxi$ and search for gauge invariant subsets of diagrams
in fully massive case;
2) to compare with the result in the unitary gauge as an internal cross-check;
3) to propose a way of realization of the step 
{\underline{from FD to renormalized}} 
{\underline{amplitudes}} within {\tt CalcPHEP} project;
4)  to compare with existing in the li\-te\-rature results;
5) to create a {\tt FORTRAN} code for the IBA (Improved Born Approximation)
cross-section
$d\sigma/dt\sim\left|A\right|^2=\left|A_{\sss BORN}+A_{\sss WEAK}\right|^2$
for subsequent use within MPT and within ``algebraic'' approach, see Section 4.
%---------------------------------------
\subsection{Amplitudes in $L,Q,D$ basis}
%---------------------------------------
In presence of massive fermions it is convenient to introduce the so-called
$L,Q,D$ basis in which the amplitude may be
parameterized with {\bf six} scalar form factors:
%---
\bqa
A^{}_{\ph} &=& \ib\frac{4\pi\qe\qf}{\sman}
~{\alpha(\sman)}~\gadu{\mu} \otimes \gadu{\mu}\,,
\nonumber
\nl
\label{Born_like}
{\cal A}^{}_{\sss{\zb}}&=&
\ib\gspi\,e^2\,4\tcie\tcit\frac{\chi_{\sss{Z}}(\sman)}{\sman}
\biggl\{\gadu{\mu}\gdp \otimes \gadu{\mu}\gdp
~{\vvertil{}{\sss{LL}}{\sman,\tman}}~
\nl [1mm] &&
-4|\qe|\siws\gadu{\mu}\otimes\gadu{\mu}\gdp
                        ~{\vvertil{}{\sss{QL}}{\sman,\tman}}
-4|\qt|\siws\gadu{\mu}\gdp\otimes\gadu{\mu}
                        ~{\vvertil{}{\sss{LQ}}{\sman,\tman}}
\nl [3mm] &&
+16|\qe\qt|\siwf\gadu{\mu}\otimes\gadu{\mu}
~{\vvertil{}{\sss{QQ}}{\sman,\tman}}~
\nl [1mm] &&
- \gadu{\mu}\gdp
\otimes i m_t D_{\mu} {\vvertil{}{\sss{LD}}{\sman,\tman}}
+ 4|\qe|\siws\gadu{\mu}\otimes i m_t D_{\mu}{\vvertil{}{\sss{QD}}{\sman,\tman}}
\biggr\}.\quad
\eqa
%---
where $\gdp=1+\gfd$,
$D_\mu = \left(p_{\bar{t}} - p_{t} \right)_\mu\,$ and
$\chi_{\sss{\zb}}(\sman)$ is the $\ph/\zb$ propagator ratio.
Every form factor in $\Rxi$ gauge could be represented as a sum of two terms:
$\vvertil{\xi}{\sss{L,Q,D}}{\sman}=\vvertil{(1)}{\sss{L,Q,D}}{\sman}
+ \vvertil{add}{\sss{L,Q,D}}{\sman,\xi}.$
First term corresponds to $\xi=1$ gauge and the second contains all $\xi$
dependences and vanishes for $\xi=1$.
We checked the cancellation of $\xi$'s separately for six subsets of diagrams: 
1) with virtual $\gamma$'s, i.e. QED subset; 2) --- $Z,\phi^0$, the so-called 
$Z$ {\em cluster}, i.e. vertices with wave function renormalization factors,
see \fig{Fig:Z_cluster};
3) --- $H,\phi^0$, i.e. $H$ cluster;
4) --- $W,\phi^{\pm}$, i.e. $W$ cluster plus all self-energies and $WW$ box;
5) four $Z\gamma$ box diagrams; 6) two $ZZ$ box diagrams.

%----------------------------------------------
\subsection{Scalar form factors of $Z$ cluster}
\begin{figure}[h!]
\vspace*{-20mm}
\[
\ba{ccccccccc}
&&
\begin{picture}(55,88)(0,41)
%--
  \Photon(0,44)(25,44){3}{5}
  \Vertex(25,44){2.5}
  \PhotonArc(0,44)(44,-28,28){3}{9}
  \Vertex(37.5,22){2.5}
  \Vertex(37.5,66){2.5}
  \ArrowLine(50,88)(37.5,66)
  \ArrowLine(37.5,66)(25,44)
  \ArrowLine(25,44)(37.5,22)
  \ArrowLine(37.5,22)(50,0)
  \Text(33,75)[lb]{$\fbf$}
  \Text(21,53)[lb]{$\fbf$}
  \Text(49,44)[lc]{$\zb$}
  \Text(21,35)[lt]{$\ff$}
  \Text(33,13)[lt]{$\ff$}
  \Text(1,1)[lb]{}
%--
\end{picture}
\quad &+&
\begin{picture}(55,88)(0,41)
%--
  \Photon(0,44)(25,44){3}{5}
  \Vertex(25,44){2.5}
  \DashCArc(0,44)(44,-28,28){3}
  \Vertex(37.5,22){2.5}
  \Vertex(37.5,66){2.5}
  \ArrowLine(50,88)(37.5,66)
  \ArrowLine(37.5,66)(25,44)
  \ArrowLine(25,44)(37.5,22)
  \ArrowLine(37.5,22)(50,0)
  \Text(33,75)[lb]{$\fbf$}
  \Text(21,53)[lb]{$\fbf$}
  \Text(47.5,44)[lc]{$\hkn$}
  \Text(21,35)[lt]{$\ff$}
  \Text(33,13)[lt]{$\ff$}
  \Text(1,1)[lb]{}
%--
\end{picture}
\; &+&
\begin{picture}(55,88)(0,41)
%--
  \Photon(0,44)(25,44){3}{5}
  \Vertex(25,44){2.5}
  \ArrowLine(50,88)(25,44)
  \ArrowLine(25,44)(50,0)
  \Text(33,75)[lb]{$\fbf$}
%  \Text(47.5,44)[lc]{$\zb$}
  \Text(33,13)[lt]{$\ff$}
  \Text(1,1)[lb]{}
\SetScale{2.0}
  \Line(17.5,17.5)(7.5,27.5)
  \Line(7.5,17.5)(17.5,27.5)
%--
\end{picture}
%--
\nl \nl  [10mm]
% second row
%--- Second row
& &
& &
\begin{picture}(75,20)(0,7.5)
%--
\Text(12.5,13)[bc]{$\ff$}
\Text(62.5,13)[bc]{$\ff$}
\Text(37.5,26)[bc]{$\ff$}
\Text(37.5,-8)[tc]{$\zb$}
\Text(0,-7)[lt]{$ $}
%--
  \ArrowLine(0,10)(25,10)
%--
  \ArrowArcn(37.5,10)(12.5,180,0)
  \PhotonArc(37.5,10)(12.5,180,0){3}{7}
%  \DashCArc(37.5,10)(12.5,180,0){3}
  \Vertex(25,10){2.5}
  \Vertex(50,10){2.5}
%--
  \ArrowLine(50,10)(75,10)
%--
\end{picture}
\; &+&
\begin{picture}(75,20)(0,7.5)
%--
\Text(12.5,13)[bc]{$\ff$}
\Text(62.5,13)[bc]{$\ff$}
\Text(37.5,26)[bc]{$\ff$}
\Text(37.5,-8)[tc]{$\hkn$}
\Text(0,-7)[lt]{$ $}
%--
  \ArrowLine(0,10)(25,10)
%--
  \ArrowArcn(37.5,10)(12.5,180,0)
  \DashCArc(37.5,10)(12.5,180,0){3}
  \Vertex(25,10){2.5}
  \Vertex(50,10){2.5}
%--
  \ArrowLine(50,10)(75,10)
%--
\end{picture}
\ea
\]
\vspace*{-1mm}
\caption{$Z$ cluster.\label{Fig:Z_cluster}}
\end{figure}
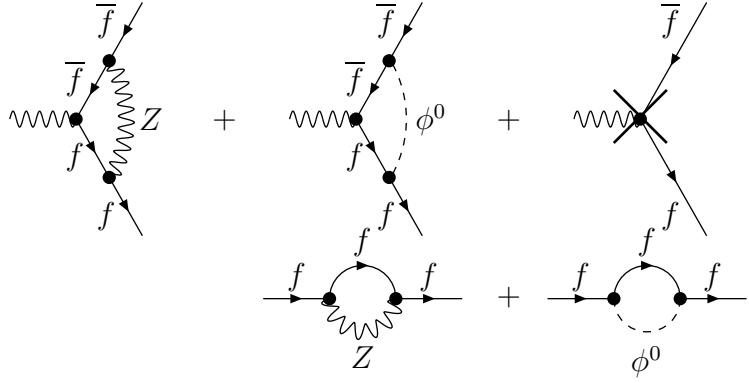

%----------------------------------------------
As an example we discuss scalar form factors of $Z$ cluster, 
\fig{Fig:Z_cluster}.
\vspace*{-1mm}

\noindent
They all are gauge invariant objects and four of them are UV-finite.
We show pole contributions ($1/{\bar\varepsilon}$) explicitly and denote 
finite quantities with $\cvertil{}{}{\sman}$:
%---
\bqa
 \vvertil{\gamma{\sss Z}}{\sss{L}}{\sman}
&\hspace{-2mm}=\hspace{-2mm}&
 \cvertil{\gamma{\sss Z}}{\sss{L}}{\sman},\,
\vvertil{\gamma{\sss Z}}{\sss{Q}}{\sman}  = 
 \cvertil{\gamma{\sss Z}}{\sss{Q}}{\sman},\,
\vvertil{\gamma{\sss Z}}{\sss{D}}{\sman}  = 
 \cvertil{\gamma{\sss Z}}{\sss{D}}{\sman},\,
\vvertil{z{\sss Z}}{\sss{D}}{\sman}       =
 \cvertil{z{\sss Z}}{\sss{D}}{\sman},
\nl[2mm]
 \vvertil{z{\sss Z}}{\sss L}{\sman} 
&\hspace{-2mm}=\hspace{-2mm}&
 -\frac{1}{4} \rtw 
\pole +\cvertil{z{\sss Z}}{\sss L}{\sman},\quad
  \vvertil{z{\sss Z}}{\sss{Q}}{\sman}= 
 -\frac{1}{16} \frac{1}{\qum \siws} \rtw \pole
 +\cvertil{z{\sss Z}}{\sss{Q}}{\sman}.
\label{poleseparation}
\eqa
%---
We present explicitly only one ``calligraphic'' quantity: 
%---
\bqa
{\cal F}^{\gamma{\sss Z}}_{\sss L} \lpar\sman\rpar 
&\hspace*{-1.35mm}=\hspace*{-1.35mm}&
 \frac{Q_t\vu}{\cows}\biggl\{
            2 \lpar 2+\frac{1}{\Rz} \rpar
       \mzs \cff{0}{-\mts}{-\mts}{-s}{\mt}{\mz}{\mt}
 -\Lmmt
\nl  &&\hspace*{-5mm}
\label{Z_cluster}
 -3\fbff{0}{-s}{\mt}{\mt}
           -2 \lpar 1+4 \ruz \rpar
                       \frac{\mzs}{4 \mts-s}
                       \Longab{0}{\mt}{\mz}  
\\  &&\hspace*{-5mm}
           + 2 \fbff{0}{-\mts}{\mt}{\mz} 
           +\frac{1}{\ruz} \Big[
               \fbff{0}{-\mts}{\mt}{\mz} + \Lmmz - 1 \Big]
\biggr\},
\nonumber
\eqa
%---
where we introduced definitions:
\bqa
&&\hspace*{-5mm}
\siws=1-\cows\,,\;\;\,
\cows=\frac{\mws}{\mzs}\,,\;\;\,
\Rz = \frac{\mzs}{s}\,,\;\;\,
 \ruz =  \frac{\mts}{\mzs}\,,\;\;\,
\ds{L_\mu(M^2)=\ln\frac{M^2}{\mu^2}}\,,
\nl[2mm] 
&&\hspace*{-5mm}
L_{ab}(M_1,M_2,M_3) =
\lpar 1+\frac{M^2_1}{M^2_3} \rpar
       {M^2_3}\cff{0}{-\mts}{-\mts}{-\sman}{M_3}{M_2}{M_3}
\nl[3mm] &&\hspace{4cm}
   - \fbff{0}{-\sman}{M_3}{M_3} + \fbff{0}{-\mts }{M_2}{M_3},
\eqa
%---
and
\vspace*{-8mm}

%--
\bq
 \bff{0}{-\sman}{M_1}{M_2}=\pole+\fbff{0}{-\sman}{M_1}{M_2}.
\eq
%--
We emphasize that we leave t'Hooft scale parameter $\tHs$ in our formulae 
unfixed, retaining an opportunity to control $\tHs$-independence 
(and therefore UV-finiteness) 
in numerical realization of one-loop form factors, providing thereby 
an additional cross-check. 

In \cite{Bardin:2000kn} we present many examples of numerics which
exhibit very good level of agreement between {\tt ZFITTER} and new code.
%----------------------------------------------
\section{``Algebraic'' approach to multi-loops}
%----------------------------------------------
There is a lot of algebraic structure in Feynman diagrams, and the idea is
to exploit it to the maximum.
That this is indeed possible was discovered in the so-called
integration-by-parts, {\em i-b-p},  
algorithm~\cite{Tkachov:1981}--\cite{Chetyrkin:1981qh}.
It proved to be hugely successful --- many well-known NNLO calculations in QCD rely 
on {\em i-b-p}.
In 1996, F.Tkachov came up with a mathematical result and a scenario to attack 
arbitrary multi-loop diagrams~\cite{Tkachov:1996}. 
It is far from being clear whether it is possible to 
use it even for simplest 2-loop diagrams. First of all, 
worth trying it for a familiar one-loop setting. 
We did it for the scalar form factors ${\cal F}^{\ph\zb}_{\sss L,Q,D}(\sman)$
of $\zb$ cluster in the process $\fep\fem\to\ft\bar{t}$, see~\fig{Fig:Z_cluster}.
Here are some results we have recently obtained~\cite{Bardin:2000cf}.
%--------------------------------------------------------------------------
\subsection{Passarino--Veltman $\scff{0}$ in variables of simplex}
%--------------------------------------------------------------------------
We begin with the usual PV function $\scff{0}$, see, 
for instance~\cite{Bardin:1999ak}: 
%---
\bqa
\ib\pi^2\,\cff{0}{\pones}{\ptwos}{\Trmoms}{\mone}{\mlone}{\mtwo}=\tHs^{\fmon} 
\intmomi{n}{\imom}\frac{1}{\dpropi{0}\dpropi{1}\dpropi{2}}\,,
\label{C0_definition}
\eqa
%---
where
%--
\bqa
&&\hspace*{-5mm}
\dpropi{0}=\imoms+\mones-\ib\ep\,,\quad
\dpropi{1}=\lpar\imom+\pone \rpar^2+\mlones-\ib\ep\,,\quad
\dpropi{2}=\lpar\imom+\Trmom\rpar^2+\mtwos-\ib\ep\,,
\nl
&&\hspace*{-5mm}
\Trmom=\pone+\ptwo,\qquad\quad\;\Trmoms=\lpar\pone+\ptwo\rpar^2=-\sman.
\eqa
%---
It is very useful to write it down in the so-called {\em simplex} variables.
After standard calculations, we arrive at an integral over two
Feynman parameters:
%---
\bqa
\cff{0}{\pones}{\ptwos}{\Trmoms}{\mone}{\mlone}{\mtwo}
=\intfx{\xvar}\int^{1-\xvar}_{0}d\yvar\,
\frac{1}{P_{\xvar,\yvar}}\,,
\label{C0_Feynman}
\eqa
%---
\vspace*{-7mm}

\noindent with
%--
\bq
\label{Pxy_polyn}
P_{\xvar,\yvar} = -\xvars\,\pones-\yvars\,\ptwos
                  +\xvar\,\yvar\bigl(\Trmoms-\pones-\ptwos\bigr) 
+\xvar\bigl(\pones+\mones-\mlones\bigr)
+\yvar\bigl(\ptwos+\mtwos-\mlones\bigr)
+\mlones.
\eq
%--
We emphasize that both the integration domain ({\em unit triangle})
and integrand are symmetric with respect to
interchange of $\xvar\leftrightarrow\yvar$ 
(and $\pone\leftrightarrow\ptwo,\;\mone\leftrightarrow\mtwo$).

Imagine now that the integral of \eqn{C0_Feynman} is incalculable analytically 
(di\-logariphms are not discovered!) and that the only way to compute it is 
a numerical integration over Feynman parameters. Representations of
\eqn{C0_Feynman} is very bad for numerical treatment due to existence of zeroes
in polynomial $P_{\xvar,\yvar}$.
%--------------------------------------------------------------------------
\subsection{``Lifting'' of polynomial powers}
%--------------------------------------------------------------------------
As was proved in~\cite{Tkachov:1996}, there exists a differential operator  
with an aid of which it is always possible to ``lift'' the power of 
a polynomial: 
%---
\bqa
 P^k_{x,y} % \lpar \mtl,\mzl,\mtl \rpar 
    &=&\frac{1}{\bf \Delta}
         \biggl[ 1-% \frac{1}{2 (k+1)}  
\frac{ \lpar \xvar + A_x \rpar \pd{x}
     + \lpar \yvar + A_y \rpar \pd{y} }{2(k+1)} \biggr]
           P^{k+1}_{x,y}.
\label{main_lifting}
\eqa
%---
Repeating procedure recursively several times, it is possible to transform any 
negative power into any positive power.
\eqn{main_lifting} involves some {\em new determinant} ${\bf \Delta}$. 
For instance, for $ P_{x,y}$ one has:
%---
\bqa
\hspace*{-1cm}
 P_{x,y}\equiv P_{x,y} \lpar \mtl,\mzl,\mtl \rpar &=& 
Q^2 \xvar\yvar+\mts \lpar \xvar+\yvar \rpar^2
                            +\mzs \lpar 1-\xvar-\yvar \rpar,
\nll 
 P_{x,y} \lpar \mtl,\mzl,\mtl \rpar &\to & 
{\bf \Delta} 
= \frac{\Delta \lpar \mtl,\mzl,\mtl \rpar}{\Delta_3}\,,
                       \quad A_{x(y)}=Q^2 
\frac{\mzs}{4\Delta_3}\,,
\eqa
%---
where\\
${\Delta_3}=-\frac{1}{4} Q^2 (Q^2+4\mts)$
is the Gram determinant for our 3-point function.\\
Vector $\Kvec{A}$ and new determinant $\Delta$ are inherent to a given diagram,
contrary to Gram determinant which depends only on external momenta,
they feel all internal masses of a diagram.
%-------------------------
\subsection{New reduction}
%-------------------------
Exploiting identities \eqn{main_lifting} and a reduction in $n$-dimensions which is
an alternative to PV reduction, we express the scalar form factors of our
$\zb$ cluster in terms of untaken integrals over Feynman parameters. 
The latter appear {\em inside} and
{\em outside} polynomials (in powers $k+\varepsilon,\;k=1,0$).
All $\xvar$ and $\yvar$, which appear outside,
may be eliminated yet in $n$-dimensions.
This is {\em a reduction} because it reduces expressions to a very limited 
number of functions which might be termed as {\em new scalars}.
Reduction in $n$-dimensions heavily exploits {\em i-b-p} and the symmetry of 
simplex.
\vspace*{-3mm}
%--------------------------
\subsubsection{New scalars}
%--------------------------
In framework of this approach we meet an analog of the usual $\scff{0}$ and 
$\sbff{0}$ functions:
%--
\bqa
{C_0 \lpar k,\tHmus; p^2_1, p^2_2,Q^2; m_1, m, m_2 \rpar}
&=& \intfx{\xvar}\int^{1-\xvar}_{0}d\yvar\,
P^k_{\xvar,\yvar}\ln\frac{P_{\xvar,\yvar}}{\tHmus}\,,
\\
{B_0 \lpar k,\tHmus; Q^2; M_1, M_2 \rpar}
&=& \Simplexl P^k_{\xvar}\ln\frac{ P_{\xvar}}{\tHmus}\,,
\eqa 
%--
where in the most general case:
$P_{\xvar}=Q^2\xvar\lpar 1-\xvar\rpar+ M^2_1\xvar+ M^2_2\lpar 1-\xvar\rpar\,.$

All new scalars, not only $\sbff{0}$, depend on the t'Hooft scale $\tHs$.
We use short hand notation for new scalars:
%---
\bqa
{\hat L}_k(\mtl,\mzl,\mtl)&=& {\mz^{-2k}}
C_{0}\lpar k,\tHmus;-\mts,-\mts,-s;\mtl,\mzl,\mtl \rpar,
\nll
{\hat L}_k(\mtl,0,\mtl)   &=& {\mz^{-2k}}
B_{0}\lpar k,\tHmus;-s;\mtl,\mtl \rpar,
\nll
{\hat L}_k(\mtl,\mzl,0)   &=& {\mz^{-2k}}
B_{0}\lpar k,\tHmus;-\mts;\mtl,\mzl \rpar,
\eqa
%---
Next, we present some results which were derived after application of 
the procedure {\em one} and {\em two times}.
\vspace*{-2mm}
%------------------------------------------
\subsubsection{``Once lifted'' expressions}
%------------------------------------------
We show only one scalar form factor $F^{\gamma Z}_{{\sss L},1}$
as a typical example:
%---
\bqa
F^{\gamma Z}_{{\sss L},1} &=&\frac{\qt\vt}{\cows}
 \biggl\{ 2{\lpttzt}-\Lnmutm+1 
-\frac{1}{\rtz} \Bigl[{\lpstzs}
\nll &&\hspace*{-8mm}
-\Lnmuzm+1 \Bigr]
- \frac{1}{\Rz}\lpar{\lpstnt} - 2{\lpttzt}-1 \rpar 
\biggr\}.\quad
\label{once}
\eqa
%---
Expression for $F^{\gamma Z}_{{\sss L},1}$ is remarkably compact, 
cf.~\eqn{Z_cluster}, and does not contain any determinants at all 
(the latter property is not typical, however).

{\bf ``Twice lifted'' expressions} have a similar structure, however, 
expressions blow up with increasing the number of recursions.

In \cite{Bardin:2000cf}
we show results of numerical computation of $F^{\gamma Z}_{\sss{L,S}}$, i.e. 
standard approach giving an analytic expressions~\eqn{Z_cluster}
in terms of dilogs; and of numerical computation of 
$F^{\gamma Z}_{\sss{L},1}$, i.e. ``once~ lifted'' expressions~\eqn{once}
and $F^{\gamma Z}_{\sss{L},2}$, i.e. ``twice lifted''.
We also show numbers for $F^{\gamma Z}_{\sss{Q},(\sss{S},1,2)}$.
Numbers show 9-10 digit agreement for Re part and 5 digits --- for Im part at 
$k=1$. This demonstrates that approach succeeds at one loop. 
%----------------
\section{Outlook}
%----------------
In my opinion, nowadays in the field of theoretical support of experiments,
it is reasonable to work in two directions:

1) Complete one-loop SM corrections for processes with number of particles 
in the final state $N_f\geq 3$
(for applications at TEVATRON, LHC, LC, $\mu$-factory).
Our project {\tt CalcPHEP} belongs to this direction.
In connection with this, one should say that 
if one wants to attack a problem of a certain level complexity, 
not starting just from the previous level, all the intermediate levels 
have to be worked through. This will considerably delay the realization of
the project. Furthermore, complete one-loop corrections for the process 
$e^+e^-\to f\bar{f}\gamma$ 
(where $f$ is any fermion, including electron) is a part of the two-loop 
program for the $Z$ resonance, so we naturally come to the second direction,
namely: 

2) Work towards two-loop precision level control of HEP observables
(for applications at GigaZ LC option).
``Algebraic'' approach for two-loops belongs to this direction.
It undoubtedly works and possesses appealing features at one loop.
Beyond one loop, there are huge algebraic difficulties;
explicit solution has never been found.
F.Tkachov currently explores scenarios that
look bizarre enough to hold promises for such a complicated problem.
Are there any chances of success in a foreseeable future?
One may count only on: 1) a non-standard F.Tkachov's expertise
which integrates a non-trivial understanding of the math
involved with a considerable skill in algorithm design and software 
engineering; and on 2)
our community intentions to continue digging in this direction, 
in particular, now in the two-loop field 
(see Giampiero Passarino's talk at this Symposium).
\vspace*{5mm}

\centerline{\large\bf\em Happy birthday, Alberto!}


\begin{thebibliography}{15}

\bibitem{fourtalks} I mean talks by T.~Kinosita, W.~Marciano,
W.~Hollik and G.~Passarino.
\vspace*{-6.2mm}

\bibitem{Bardin:1999yd}
D.~Bardin, M.~Bilenky, P.~Christova, M.~Jack, L.~Kalinovskaya, A.~Olchevski, 
S.~Riemann, and T.~Riemann, 
  {\em Comput. Phys. Commun.} {\bf 133} (2000) 229--395, {\tt hep-ph/9908433}.
%``ZFITTER v.6.21: A semi-analytical program for fermion pair
%  production in $e^{+}e^{-}$ annihilation'', DESY--Zeuthen preprint 99-070
%  (1999), 
\vspace*{-1.6mm}

\bibitem{Bardin:1999ak}
D.~Bardin and G.~Passarino, ``The standard model in the making: Precision study
  of the electroweak interactions'', Oxford, UK: Clarendon, 1999.
\vspace*{-1.6mm}

\bibitem{CalcPHEP:2000}
G.~Passarino, D.~Bardin, L.~Kalinovskaya, P.~Christova, G.~Nanava, A.~Andonov,
  and S.~Bondarenko, {\tt CalcPHEP}, a write-up in preparation.
\vspace*{-1.6mm}

\bibitem{Hahn:2000jm}
T.~Hahn, {\em Nucl. Phys. Proc. Suppl.} {\bf 89} (2000) 231--236.
\vspace*{-1.6mm}

\bibitem{Beneke:2000hk}
M.~Beneke {\em et~al.}, ``Top quark physics'', in {\em Proc. of the Workshop on
  Standard Model Physics (and more) at the LHC, CERN 2000--004} (G.~Altarelli
  and M.~Mangano, eds.), pp.~419--529, 2000.
\vspace*{-1.6mm}

\bibitem{Denner:2000bj}
A.~Denner, S.~Dittmaier, M.~Roth, and D.~Wackeroth, {\em Nucl. Phys.} {\bf
  B587} (2000) 67--117.
\vspace*{-1.6mm}

\bibitem{Nekrasov:1999cj}
M.~L. Nekrasov, ``Finite width effects and gauge cancellations in $\wb$ and
  $\zb$ boson production in framework of modified perturbation theory'', 
{\tt hep-ph/0002184}, {\em Eur.Phys.J.} {\bf C19} (2001) 441--454 .
\vspace*{-1.6mm}

\bibitem{Bardin:2000kn}
D.Bardin, L. Kalinovskaya and G. Nanava,
``An electroweak library for the calculation of EWRC to 
$e^+e^-\to f{\bar f}$ within the {\tt CalcPHEP} project'',
{\tt hep-ph/0012080}, revised version, November 2001, CERN-TH/2001-308. 
\vspace*{-1.6mm}

\bibitem{Beenakker:1991ca}
W.~Beenakker, S.~C. van~der Marck, and W.~Hollik, {\em Nucl. Phys.} {\bf B365}
  (1991) 24--78.
\vspace*{-1.6mm}

\bibitem{Driesen:1996tn}
V.~Driesen, W.~Hollik, and A.~Kraft, ``Top pair production in $e^{+}e^{-}$
  collisions with virtual and real electroweak radiative corrections'',
  Karlsruhe University preprint (1996), {\tt hep-ph/9603398}.
\vspace*{-1.6mm}

\bibitem{Tkachov:1981}
F. Tkachov, {\em Phys. Lett.} {\bf 100B} (1981) 65--68.
\vspace*{-1.6mm}

\bibitem{Chetyrkin:1981qh}
K.G. Chetyrkin and F.V. Tkachov, {\em Nucl. Phys.} {\bf B192} (1981) 159--204.
\vspace*{-1.6mm}

\bibitem{Tkachov:1996}
F. Tkachov, {\it NIM} {\bf A389} (1997) 309--313.
\vspace*{-1.6mm}

\bibitem{Bardin:2000cf}
D.Bardin, L. Kalinovskaya and F. Tkachov,
``New algebraic-numeric methods for loop integrals: Some 1-loop experience'',
{\tt hep-ph/0012209}.
\vspace*{-10mm}

\end{thebibliography}
\end{document}